\pdfoutput=1
\documentclass[journal,twocolumn]{IEEEtran}
\usepackage{cite}
\usepackage{graphicx}
\usepackage[cmex10]{amsmath}
\usepackage{color}
\newtheorem{proposition}{\bf   Proposition}
\newtheorem{theorem}{\bf Theorem}

\begin{document}
%
\title{Competitive Resource Allocation in HetNets: the Impact of Small-cell Spectrum Constraints and Investment Costs }
%
%
%

\author{Cheng~Chen,~\IEEEmembership{Member,~IEEE,}
        Randall~A.~Berry,~\IEEEmembership{Fellow,~IEEE,}
        Michael~L.~Honig,~\IEEEmembership{Fellow,~IEEE,}
        and~Vijay~G.~Subramanian,~\IEEEmembership{Member,~IEEE}
\thanks{Cheng Chen was with the Department of Electrical Engineering and Computer Science,
Northwestern University, Evanston, IL. He is now with the Next Generation and Standards
Group at Intel Corporation, Hillsboro, OR. Email: cheng.chen@intel.com.}

\thanks{Randall A. Berry and Michael L. Honig are with the Department
of Electrical Engineering and Computer Science, Northwestern University, Evanston,
IL. Emails:  \{rberry, mh\}@eecs.northwestern.edu.}
\thanks{Vijay G. Subramanian is with the department of Electrical Engineering and Computer Science, University of Michigan, Ann Arbor, MI. Email: vgsubram@umich.edu.}
\thanks{This work was presented in part at 2016 the 50th Annual Conference on Information Sciences and Systems, Princeton, NJ \cite{Investment3-Chen16} and 2017 IEEE International Symposium on Dynamic Spectrum Access Networks, Baltimore, MD \cite{Policy5-Chen17}.}
\thanks{This research is supported in part by NSF grant 1343381.}}
\maketitle

\begin{abstract}
Heterogeneous wireless networks with small-cell deployments in licensed
and unlicensed spectrum bands are a promising approach for
expanding wireless connectivity and service.
As a result, wireless service providers (SPs) are adding small-cells
to augment their existing macro-cell deployments. This added flexibility
complicates network management, in particular, service pricing and spectrum allocations
across macro- and small-cells. Further, these decisions depend on the degree of competition
among SPs.  Restrictions on shared spectrum access imposed by regulators,
such as low power constraints that lead to small-cell deployments,
along with the investment cost needed to add small cells to an existing network,
also impact strategic decisions and market efficiency.
If the revenue generated by small-cells does not cover the investment cost,
then there will be no deployment even if it increases social welfare.
We study the implications of such spectrum constraints and investment costs
on resource allocation and pricing decisions by competitive SPs,
along with the associated social welfare.
Our results show that while the optimal resource allocation
taking constraints and investment into account can be uniquely determined,
adding those features with strategic SPs can have a substantial effect
on the equilibrium market structure.
\end{abstract}


\section{Introduction} \label{Sec:Introduction}
It is generally expected that current cellular networks will continue to evolve towards heterogeneous networks (HetNets) to accommodate the explosive demand for wireless data \cite{HetNet1-Qualcomm11,HetNet2-Ghosh12}. This will require service providers (SPs) to increasingly deploy small-cells in addition to traditional macro-cells. Macro-cells, which typically have large transmission power,  are capable of covering users within a large region. In contrast, small-cells have much lower transmission power and are used to provide service to a local area.

While the deployment of small-cells will increase overall data capacity, it also makes the network management and resource allocation more complex for SPs that are (competitively) operating their HetNets. This includes price differentiation and optimal splitting of their limited bandwidth resources between macro- and small-cells. These decisions must also take into account the fact that users in the network are also heterogeneous, in terms of their mobility. In this paper, we study such issues and in particular we concentrate on two important factors that can impact these decisions: small-cell spectrum restrictions and investment choices. By ``spectrum restrictions", we are refering to restrictions that certain bands of spectrum can only be used for small-cell deployment. This is motivated by the FCC policy for the new Citizens Broadband Radio Service
(CBRS) in the 3550-3650 MHz band (3.5GHz Band) \cite{FCC}. Restricting usage of this band to small-cells, enables stricter power regulations, which in turn helps facilitate sharing the band with existing incumbent users. Such restrictions will clearly impact a SP's network management decisions both within the restricted band and in other bands. The amount of small-cells deployed also will depend on the investment decisions made by SPs for deploying this new infrastructure.
In a competitive environment, these decisions will be coupled among different SPs and will also depend on the SPs existing sunk investment in macro-cells.

Our approach in this paper builds on prior work in \cite{Opt7-Chen13,Competition5-Chen15}
that developed a model for studying pricing and bandwidth allocation decisions among competitive SPs in HetNets. In this model, users decide on the rate to request from a SP by maximizing the difference between the utility they receive and the cost of service. A key result in \cite{Opt7-Chen13,Competition5-Chen15} is that for the class of $\alpha$-fair utility functions, the Nash equilibrium achieved by revenue optimizing SPs achieves the optimal social welfare. This prior work did not consider bandwidth restrictions or investment decisions, i.e., SPs could use their bandwidth for either small- or macro-cells and any investment infrastructure was assumed to already be sunk. Here, we consider both of these effects which results in non-trivial generalizations of the work in \cite{Opt7-Chen13,Competition5-Chen15} and leads to significantly different conclusions.  For example, we show that either of these considerations can lead to a loss in social welfare with $\alpha$-fair utilities. We analyze both a single monopolist SP and a competitive scenario with multiple SPs.

We first study the influence of spectrum restrictions. Here we assume the SPs have the same predetermined infrastructure deployment densities and can't change these by investing in additional infrastructure. The spectrum restriction considered is such that each SP has a minimum amount of bandwidth that can only be allocated to small-cells. Then we pivot to studying the impact of investment and assume the SPs can partition the bandwidth freely, subject to no restrictions. However, there is a per unit deployment cost of small-cells, modeling the investment needed by the SPs. In contrast, we assume that any investment in macro-cells has already been sunk. In both of these cases we characterize the SPs optimal bandwidth partition between small- and macro-cells and their optimal pricing decisions. Finally, we evaluate the impact of the two factors on the social welfare achieved.

\subsection{Contributions}

We now summarize our primary contributions in this paper:

1. \emph{Incorporating spectrum restrictions or small-cell investment into the HetNet Model}:
As noted prior related work did not incorporate either of these concerns.  Here, we give a models that
incorporate each consideration.

2. \emph{Characterizing the impact of spectrum restrictions and investment costs for SPs}: We analyze scenarios with both a monopoly SP and competitive SPs, and illustrate the impacts brought by introducing spectrum restrictions and the investment in small-cells. We show that with small-cell spectrum restrictions, a monopolist SP will simply increases its small-cell bandwidth to the required minimum amount if its small-cell bandwidth without restrictions is less than the constraint. This applies to both social welfare and revenue-maximization. With two competitive SPs, there always exists a unique Nash equilibrium that depends on the spectrum restriction.  We illustrate this by considering three cases corresponding to whether the equilibrium allocation without restrictions satisfies the two constraints. We characterize the equilibrium for each case.
In contrast, with investment costs, the monopoly SP only invests in small-cells if the per unit deployment cost for small-cells is below a threshold. In a competitive scenario, since a general analysis appears to be difficult, we focus on a simplified model with two SPs and a binary investment choice in which the small-cell deployment density is fixed.
We show that depending on the associated deployment costs, different types of Nash equilibrium are possible corresponding to different scenarios where one, none, or both SPs invest.


3. \emph{Social Welfare Analysis}: We characterize the social welfare for both spectrum constraints and investment costs. For spectrum constraints, we show that if the equilibrium without constraints violates the constraints, then social welfare loss is inevitable. However, the social welfare loss is always bounded, and the worst case happens when the spectrum regulator requires the SPs to allocate all bandwidth only to small-cells. With investment, we
show that a monopoly SP, if it does invest in small-cells, should deploy a higher density of small-cells to maximize welfare than when maximizing revenue. For two competitive SPs with a binary investment choice, again we show that the Nash equilibria may not be socially optimal.

\subsection{Related Work}
Pricing and bandwidth allocation problems in HetNets have attracted considerable attention. In \cite{Opt1-Shetty09, Opt2-Gussen11, Opt3-Yun11}, small-cell service is considered as an enhancement to macro-cell service . In contrast, \cite{Opt4-Chen11, Opt5-Lin11, Opt6-Duan13} consider macro-cell and small-cell service as separate services, the same as in this paper. Only optimal pricing is studied in \cite{Opt1-Shetty09, Opt3-Yun11, Competition1-Zhang13, Competition2-Hossain08}, while \cite{Competition5-Chen15,Opt2-Gussen11, Opt4-Chen11,Opt5-Lin11,Opt6-Duan13,Unlicensed1-Chen16} consider joint pricing and bandwidth allocation, as in this paper. Additionally, except for \cite{Competition5-Chen15,Competition1-Zhang13, Competition2-Hossain08,Unlicensed1-Chen16} that include  competitive scenarios with multiple SPs, all the other work assumes only one SP. In this paper, we investigate both monopoly and competitive scenarios.

The spectrum regulations in the 3.5GHz band have attracted some attention in the research literature, and are
seen as a great opportunity for small-cell networks which could enhance existing macro-cell service \cite{Policy1-Nakamura13}. However, most of the existing work focuses on the technical challenges of deploying small-cells in the 3.5 GHz band, such as path loss validation \cite{Policy2-Nguyen13}, or how to cope with the interference coming from shipborne radar systems \cite{Policy3-Yang15, Policy4-Mo16}. None of the preceding work has looked at how this band will impact bandwidth allocation and pricing in other bands as we do here. Investment costs have  largely been neglected in the preceeding work on HetNets. In \cite{Opt6-Duan13} femtocell operational cost is considered and it is linear with femtocell bandwidth. In contrast, \cite{Investment1-Markendahl10, Investment2-Frias12} take both deployment cost and operational cost into account and conduct an economic analysis with case studies or simulations.
\\

The rest of the paper is organized as follows. We present the system model in Section \ref{Sec:System Model}. We consider the impacts on small-cell resource allocation brought by introducing spectrum restrictions and small-cell investment cost separately  in Section \ref{Sec:Regulations} and Section \ref{Sec:Investment}, respectively. Social welfare analysis is in Section \ref{Sec:SW}. We conclude in Section \ref{Sec:Conclusions}. Due to space considerations, all proofs of the main results can be found in the appendix.

\section{System Model}\label{Sec:System Model}
\begin{figure}[htbp]
\centering
\includegraphics[width=0.45\textwidth,height=0.35\textwidth]{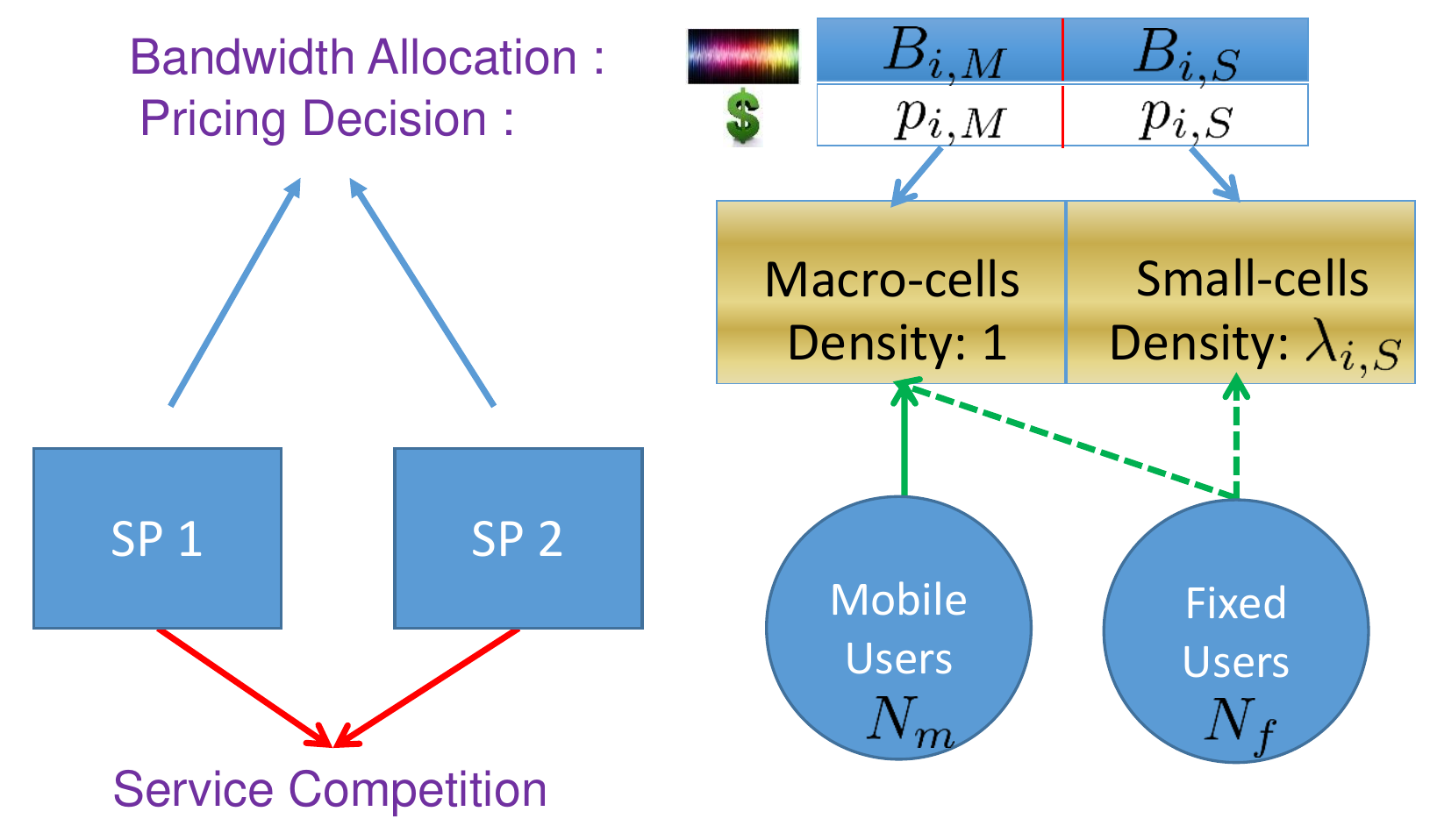}
\caption{A depiction of our overall systems model when there are 2 SPs in the market.}
\label{Fig:TCCN_System Model}
\end{figure}

We adopt a similar mathematical model as in our previous work \cite{Opt7-Chen13, Competition5-Chen15} for the analysis. Fig. \ref{Fig:TCCN_System Model} illustrates the network and market model. We now describe the different aspects of it while pointing out the additional elements considered here.

\subsection{SPs}
We consider a HetNet with $N$ SPs providing separate macro- and small-cell service to all users. Denote the set of SPs as $\mathcal{N}$. Each SP operates a two-tier cellular network consisting of macro-cells and small-cells, which are assigned different licensed bands and are deployed uniformly over a given area. We assume all SPs have the same macro-cell infrastructure density, normalized to one. In contrast, the deployment density of small-cells  of SP $i$, is denoted as $\lambda_{i,S}$.  In our setting, macro-cells have high transmission power, and therefore can provide large coverage range. In contrast, small-cells have low transmission power, and consequently local coverage range.

Each SP $i$ has a total amount of bandwidth $B_i$ exclusively licensed.\footnote{For the monopoly SP scenario, we will ignore the subscript.} Since we assume all macro- and small-cells use separate bands, each SP $i$ needs to decide how to split its bandwidth into $B_{i,M}$, bandwidth allocated to macro-cells, and $B_{i,S}$, bandwidth allocated to small-cells. When determining this partition, every SP is required to conform to (possible) bandwidth regulations enforced by the spectrum regulator. Specifically, SP $i$ is requested to guarantee a minimum amount of bandwidth allocated to small-cells, and this lower bound is denoted as $B_{i,S}^0$.

We assume that macro- and small-cells use the same transmission technology, and so have the same (average) spectral efficiency $R_0$\footnote{Recent studies show that this may not be true, and the spectral efficiency in small-cells may be 2-3x higher compared to macro-cells. In this case, we can add another spectral efficiency gain factor $\lambda_{\text{se}}$ in small-cells and the analysis still applies.}. Each SP $i$ has a total bandwidth $B_i$, which is split into $B_{i,M}$ and $B_{i,S}$, the bandwidths allocated to macro- and small-cells, respectively. Therefore, for a fixed bandwidth allocation, the total available rates provided by the macro- and small-cells for SP $i$ are $C_{i, M}=B_{i, M}R_0$ and $C_{i,S}=\lambda_{i,S} B_{i, S}R_0$, respectively. Of course, in practice, the spectral efficiency might vary with the particular spectrum used as well. We ignore such effect here. Each SP $i$ provides separate macro- and small-cell services and charges the users a price \emph{per unit rate} for associating with its macro-cells or small-cells, namely, $p_{i, M}$ and $p_{i, S}$.

\subsection{Users}
We assume the users in the networks are also heterogeneous and categorize them into two types based on their mobility patterns. Mobile users can only be served by macro-cells. In contrast, fixed users are relatively stationary, and can connect to either macro- or small-cells (but not both). Denote the densities of mobile users and fixed users as $N_m$ and $N_f$, respectively. Note that the heterogeneity of the users can also arise from an equivalent model that assumes $(N_m+N_f)$ as the total density of users, who are mobile with probability $N_m/(N_m+N_f)$ and stationary with probability $N_f/(N_m+N_f)$. After user association, let $K_{i, M}$ and $K_{i, S}$ denote the mass of users connected to the macro- and small-cells of SP $i$, respectively. (Note that $K_{i, S}$ consists of fixed users only, whereas $K_{i, M}$ can consist of both mobile and fixed users.)

\subsection{Small-cell Spectrum Restrictions}
In Section \ref{Sec:Regulations},  we consider small-cell spectrum restrictions. These are modeled by restricting the bandwidth partition of each SP to satisfy a specified minimum amount for small-cells (representing the amount of restricted spectrum that SP owns).  Specifically, SP $i$ is required to guarantee $B_{i,S} \geq B_{i,s}^0$, where $B_{i,S}^0$ is the minimum small cell bandwidth.  As indicated in the introduction, this is primarily motivated by the case of 3.5 GHz band.

\subsection{Small-cell Investment Costs}
In Section \ref{Sec:Investment}, we study the impact of small-cell investment costs associated with deploying small-cell infrastructure, modeled by a cost per unit density denoted by $I_S$. That is, if SP $i$ wants to deploy and maintain a small-cell network with $\lambda_{i,S}$ density of small-cells, it has to pay an investment cost of $\lambda_{i,S}I_S$, which has to be taken into account when the SP calculates its net operating revenue by providing both macro- and small-cell service. In Section \ref{Sec:Regulations}, we assume that any such costs are already sunk and the SPs simply have a given density of small cells.

\subsection{User and SP Optimization}
We now introduce the optimization problems corresponding to both users and SPs. We assume each user is endowed with a utility function, $u(r)$, which only depends on the service rate it gets. For simplicity, we assume that all users have the same $\alpha$-fair utility functions \cite{MoWalrand} with $\alpha \in (0,1)$:\footnote{One future direction is to relax this constraint, and allow different users to have different $\alpha$-fair utility functions to create another user heterogeneity aside from being mobile or fixed.}
\begin{equation}\label{Eqn:UtilityFn}
u(r)=\frac{r^{1-\alpha}}{1-\alpha}, \quad \alpha \in (0,1).
\end{equation}
This restriction enables us to explicitly calculate many equilibrium quantities, which appears to be difficult for more general classes of utility. Furthermore, this class is widely used in both networking and economics, where it is a subset of the class of iso-elastic utility functions.\footnote{In general $\alpha$-fair utilities require that $\alpha \geq 0$ to ensure concavity; requiring $\alpha>0$ ensures strict concavity but allows us to approach the linear case as $\alpha \rightarrow 0$.  The restriction of $\alpha<1$  ensures that utility is non-negative so that a user can always ``opt out" and receive zero utility. Note also that as $\alpha \rightarrow 1$, we approach the  $\log(\cdot)$ (proportional fair) utility function.}

Each user chooses the service rate by maximizing its net payoff $W$,
defined as its utility less the service cost. For a service with price $p$, this is equivalent to:
\begin{equation} \label{Opt:User Optimization}
W= \max_{r\geq 0}\quad u(r)-pr.
\end{equation}
For $\alpha$-fair utility functions,  (\ref{Opt:User Optimization}) has the  unique solution:
\begin{equation}\label{Eqn:UserRateOpt}
r^{*}= D(p)=(u^{\prime})^{-1}(p)=(1/p)^{1/\alpha}, 
\end{equation}
where $D(p)$ here can be seen as the user's rate demand function. The maximum net payoff for a user is thus:
\begin{equation} \label{Eqn:User Net Payoff}
W^*(p) = u(D(p))- pD(p)=\frac{\alpha}{1-\alpha}p^{1-\frac{1}{\alpha}}.
\end{equation}
Recall, that fixed users can choose between any macro- or small-cell service offered by any SP, while mobile users can only choose the macro-cell service provided by a SP. However, here, we assume mobile users have priority connecting to macro-cells, which means macro-cells will only admit fixed users after the service requests of all mobile users have been addressed.

For the association rules, we adopt the same process described in \cite{Competition5-Chen15}. That is, users always choose the service with lowest price and fill the corresponding capacity. If multiple services have the same price, then the users are allocated across them in proportion
to the capacities. Once a particular service capacity is exhausted, then the leftover demand continues to fill the remaining service in the same fashion.

Each SP determines the bandwidth split and service prices to maximize its revenue, which is the aggregate amount paid by all users associating with their macro- and small-cells, less possible investment cost. Meanwhile, with small-cell spectrum restrictions, they also need to conform to the constraints on small-cell bandwidth allocation.

Specifically, in Section \ref{Sec:Regulations}, where we study the regulatory constraints on small-cell bandwidth allocation, we assume all SPs have the same predetermined small-cell deployment density, denoted as $\lambda_S$. SP $i$ thus solves the following optimization problem:
\begin{subequations}
\begin{align}
\text{maximize}  \quad &S_i=p_{i, M}K_{i, M}D(p_{i, M})+p_{i, S}K_{i, S}D(p_{i, S}),  \\
\text{subject to}  \quad& B_{i, M}+B_{i, S}\le B_i,  B_{i, M}\ge 0, B_{i, S}\ge B_{i,S}^0,  \label {Opt:SP's optimization-Constraint4}\\
&K_{i, M}D(p_{i, M})\le C_{i,M}, K_{i, S}D(p_{i, S})\le C_{i,S}, \label {Opt:SP's optimization-Constraint5}\\
 & 0<p_{i, M}, p_{i, S}<\infty. \label {Opt:SP's optimization-Constraint6}
\end{align}
\end{subequations}

In Section \ref{Sec:Investment}, where we consider investment costs (but no spectrum restrictions), SP $i$ solves the following optimization problem:
\begin{subequations}
\begin{align}
\notag \text{maximize} \quad & S_i =  K_{i, M}p_{i, M}D(p_{i, M})+K_{i, S}p_{i, S}D(p_{i, S}) \\
 & ~~~~~ - \lambda_{i,S}I_S, \label{Opt:SP Optimization with investment}\\
&  B_{i, M}, B_{i, S}\ge 0, B_{i, M}+B_{i, S} \le B_i, \label {Opt:SP's optimization-Constraint1}\\
&K_{i, M}D(p_{i, M})\le C_{i,M}, K_{i, S}D(p_{i, S})\le C_{i,S}, \label {Opt:SP's optimization-Constraint2}\\
&0\le p_{i, M}, p_{i, S}<\infty, \quad \lambda_{i,S}\ge 0. \label {Opt:SP's optimization-Constraint3}
\end{align}
\end{subequations}

Alternatively, a social planner, such as the FCC, may seek to allocate bandwidth and set prices to maximize social welfare, which is the sum utility of all users, less possible investment costs in small-cells. With spectrum restrictions, this is given by:
\begin{align}
\text{maximize}  \quad &\text{SW}=\sum\limits_{i=1}^{N}[K_{i, M}u(D(p_{i, M}))+K_{i, S}u(D(p_{i, S}))],\label{Opt:SW optimization with regulations}
\end{align}
subject to the same constraints (\ref{Opt:SP's optimization-Constraint4}), (\ref{Opt:SP's optimization-Constraint5}) and (\ref{Opt:SP's optimization-Constraint6}).

In contrast, with investment cost, this is equivalent to:
\begin{align}
\notag \text{maximize}  \quad \text{SW}=\sum\limits_{i=1}^{N}[K_{i, M}u(D(p_{i, M}))&+K_{i, S}u(D(p_{i, S}))\\
&-\lambda_{i,S}I_S],\label{Opt:SW optimization with investment}
\end{align}
subject to the same constraints (\ref{Opt:SP's optimization-Constraint1}), (\ref{Opt:SP's optimization-Constraint2}) and (\ref{Opt:SP's optimization-Constraint3}).

\subsection{Sequential Game and Backward Induction}
We model the investment, bandwidth and price adjustments of SPs in the network as a three-stage process:
\begin{enumerate}
\item
The SPs determine their investment levels, i.e., their small-cell deployment density.
\item
Each SP $i$ first determines its bandwidth allocation $B_{i, M}, B_{i, S}$ between macro-cells and small-cells. Denote the aggregate bandwidth allocation profile as $\mathbf{B}$.
\item
Given $\mathbf{B}$ (assumed known to all SPs), the SPs announce prices for both macro-cells and small-cells. The users then associate with SPs according to the previous user association rule.
\end{enumerate}

This process is motivated by the expectation that in practice bandwidth allocation and investment take place over a slower time-scale than price adjustments. Also, we assume the bandwidth allocation happens on a faster time-scale than the investment decision, which is reasonable if a SP can dynamically change its bandwidth assignment after it has deployed its small-cells. Moreover, changing the bandwidth partition could conceivably involve reconfiguring equipment at both base stations and handsets, and adjusting the placement of access points along with transmission parameters in order to keep the rate per cell fixed. Adjustment of prices would not require these additional changes. As a result, we generally assume price adjustment happens at a faster time-scale than bandwidth allocations.\footnote{In some scenarios where the equipment is capable of operating over a wide range of frequencies without big configuration adjustments, bandwidth partition changes may also happen dynamically. If we assume the bandwidth allocation happens at a faster-time scale than price adjustments, the analysis would again use the sub-game perfect equilibrium concept, and also backward induction (with order reversed) to determine equilibria. This is out of scope of this paper.}

We then do backward induction. That is, we first derive the price equilibrium under a fixed bandwidth allocation. We then characterize the bandwidth allocation equilibrium based on the price equilibrium obtained. Finally, when small-cell investment cost is considered in Section \ref{Sec:Investment}, we compute the investment equilibrium. For the first two steps of computing the price and bandwidth allocation equilibrium, we will apply the results obtained in \cite{Opt7-Chen13} and \cite{Competition5-Chen15} to simplify the analysis.

\section{The Impact of Small-Cell Spectrum Restrictions}\label{Sec:Regulations}
In this section we investigate the impacts of regulatory constraints on small-cell bandwidth allocation on the pricing and spectrum allocation in HetNets, with the corresponding optimization problem in (5). Note that in order to evaluate the impacts of spectrum restrictions independently, and also for simplicity of analysis, in this section we assume all SPs have already made the investment in small-cells and deployed the necessary infrastructures.\footnote{Otherwise it makes no sense to enforce the constraint that each SP has to set aside a specific amount of bandwidth for small-cells.} Moreover, we further assume all SPs have deployed the same density of small-cells, denoted as $\lambda_S$. We first analyze the monopoly scenario with a single SP and show that in the monopoly scenario the SP simply increases its small-cell bandwidth to the required minimum amount if its optimal small-cell bandwidth without restrictions is less than the constraint. With two competitive SPs, and there always exists a unique Nash equilibrium that depends on the regulatory constraints. We illustrate this by considering three cases corresponding to whether the equilibrium allocation without regulatory restrictions satisfies the two constraints, and characterize the equilibrium for each case.

\subsection{Monopoly Scenario}
We first study the bandwidth allocation when a single SP is operating in the network. This is similar to the analysis in our previous work \cite{Opt7-Chen13}, except here we have an additional regulatory constraint that imposes a minimum bandwidth allocation to small-cells. This added constraint will change the optimal bandwidth allocation strategy for the monopoly SP. In \cite{Opt7-Chen13} it is concluded that for the set of $\alpha$-fair utility functions we use in this paper, the revenue-maximizing and social welfare-maximizing bandwidth allocation turn out to be the same. The following theorem states that the optimal bandwidth allocations under both objectives are still the same, but adding a large value for the bandwidth set aside for small-cells changes the optimal bandwidth allocation.

\begin{theorem}
\label{Thm:Monopoly Bandwidth Allocation}
For a monopoly SP, the optimal revenue-maximizing bandwidth allocation strategies are the same as the welfare maximizing strategies and can be determined by the following cases:

1. If $B_S^0\le \frac{N_f\lambda_S^{1/\alpha-1}B}{N_f\lambda_S^{1/\alpha-1}+N_m}$, the optimal bandwidth allocation remains the same as that without the regulatory constraint. In this case it is given by:
\begin{subequations}
\begin{align}
&B_S^{\text{SW}}=B_S^{\text{rev}}=\frac{N_f\lambda_S^{1/\alpha-1}B}{N_f\lambda_S^{1/\alpha-1}+N_m}, \\ &B_M^{\text{SW}}=B_M^{\text{rev}}=\frac{N_m B}{N_f\lambda_S^{1/\alpha-1}+N_m}.
\end{align}
\end{subequations}

2. If $B_S^0> \frac{N_f\lambda_S^{1/\alpha-1}B}{N_f\lambda_S^{1/\alpha-1}+N_m}$,  the optimal bandwidth allocation is changed to:
\begin{equation}
B_S^{\text{SW}}=B_S^{\text{rev}}=B_S^0, \quad B_M^{\text{SW}}=B_M^{\text{rev}}=B-B_S^0.
\end{equation}
Consequently there will be both a welfare and revenue loss if this case applies.

In both cases the optimal macro- and small-service prices are market-clearing prices, i.e., the prices that equalize the total rate demand and the total rate supply in both cells. (See Appendix \ref{Appen:A} for proof.)
\end{theorem}

Theorem \ref{Thm:Monopoly Bandwidth Allocation} states that if the original optimal bandwidth allocation without the bandwidth restrictions already satisfies the imposed constraint, then the SP just keeps the same bandwidth allocation. If the original bandwidth allocation violates the regulatory constraint, then the SP increases the small-cell bandwidth to the required level. This is because the added regulatory constraint does not change the concavity of the revenue or social welfare function with respect to the small-cell bandwidth, and further increasing the bandwidth allocation to small-cells will only lead to more revenue loss.

\subsection{Competitive Scenario}
We now turn to the competitive scenario with two SPs, each of which maximizes its individual revenue. Applying the results from \cite{Competition5-Chen15}, the price equilibrium given any fixed bandwidth allocation is always the market-clearing price. We therefore focus on the bandwidth allocation Nash equilibrium.

Considering the case without the additional regulatory constraint, using the results from \cite{Competition5-Chen15}, there exists a unique Nash equilibrium and the bandwidth allocations of two SPs at equilibrium are given by:
\begin{subequations}
\begin{align}
&B_{1,S}^{\text{NE}}=\frac{N_f\lambda_S^{1/\alpha-1}B_1}{N_f\lambda_S^{1/\alpha-1}+N_m}, B_{1,M}^{\text{NE}}=\frac{N_m B_1}{N_f\lambda_S^{1/\alpha-1}+N_m},\\
&B_{2,S}^{\text{NE}}=\frac{N_f\lambda_S^{1/\alpha-1}B_2}{N_f\lambda_S^{1/\alpha-1}+N_m},  B_{2,M}^{\text{NE}}=\frac{N_m B_2}{N_f\lambda_S^{1/\alpha-1}+N_m}.
\end{align}
\end{subequations}

With the additional regulatory constraints, we have the following theorem characterizing the corresponding Nash equilibrium between two SPs.

\begin{theorem} \label{Thm:NE}
With two SPs, and a constraint on minimum small-cell bandwidth, a unique Nash equilibrium exists. Moreover, the total bandwidth allocated to small-cells by the two SPs is no less than that without the regulatory constraints. See Appendix \ref{Appen:B} for proof.)
\end{theorem}

Theorem \ref{Thm:NE} states that the existence and uniqueness of the Nash equilibrium is preserved after adding the regulatory constraints. This can be proved using similar methods as provided in
\cite{Competition5-Chen15}, with some  modifications. The last part of the theorem is subtler than it appears. One may try to argue that if any of the constraints is violated, that SP then needs to increase its bandwidth allocation to small-cells. It would then hold that the total bandwidth allocated to small-cells surely increases. However, the logic does not carry through if only one constraint is violated at the Nash equilibrium omitting the constraint. In that case, the SP with violated constraint must increase the bandwidth allocation to small-cells. However, the other SP, whose equilibrium small-cell bandwidth allocation without restrictions satisfies the constraint, may potentially \emph{decrease} its bandwidth in small-cells in response to the increase in bandwidth allocation of its competitor. In that case, determining the change in total bandwidth requires a more detailed analysis. Nonetheless, Theorem \ref{Thm:NE} indicates that even here the total bandwidth in small-cells would not decrease. We will present a specific example later.

Depending on whether the regulatory constraints are violated or not at the Nash equilibrium without the constraints, there are three cases we need to cover independently. We will see that, in each case, the Nash equilibrium behaves differently.

{\it Case A: Both constraints are satisfied.}
The new Nash equilibrium is the same as the Nash equilibrium without restrictions.

{\it Case B: Both constraints are violated.}
The Nash equilibrium without restrictions is no longer valid. The following proposition characterizes the properties of the new Nash equilibrium.
\begin{proposition}\label{Prop:Both Violated NE}
In case B, the Nash equilibrium with regulatory constraints is one of the following types:\\
Type I: Both SPs increase their small-cell bandwidth allocations to exactly the required amount, i.e., $B_{1,S}=B_{1,S}^0, B_{2,S}=B_{2,S}^0$.\\
Type II: One SP increases its small-cell bandwidth exactly to the required amount, while the other SP increases further beyond the required amount, i.e., $B_{1,S}=B_{1,S}^0, B_{2,S}>B_{2,S}^0 \text{ or } B_{1,S}>B_{1,S}^0, B_{2,S}=B_{2,S}^0$.
\end{proposition}

It is conceptually easy to characterize the necessary and sufficient conditions for the first type of Nash equilibrium to hold since at that equilibrium the marginal revenue increase with respect to per unit of bandwidth increase in small-cells should be non-positive for both SPs. This can be analytically expressed via the two corresponding inequalities:
\begin{align}\label{Eqn:Boundary Point NE}
\nonumber &\lambda_S{R_S^0}^{-\alpha}-{R_M^0}^{-\alpha}-\frac{\alpha \lambda_S^2B_{i,S}^0R_0}{N_f}{R_S^0}^{-\alpha-1}+\\
&\frac{(B_i-B_{i,S}^0)R_0}{N_m}{R_M^0}^{-\alpha-1}
\le 0, \text{ for } i=1,2.
\end{align}
Here, $R_S^0$ and $R_M^0$ are defined as follows:
\begin{subequations}
\begin{align}
&R_S^0=\frac{\lambda_S(B_{1,S}^0+B_{2,S}^0)R_0}{N_f},\\
&R_M^0=\frac{(B_1-B_{1,S}^0+B_2-B_{2,S}^0)R_0}{N_m}.
\end{align}
\end{subequations}
%

{\it Case C: Only one constraint is violated.}
Without loss of generality, we assume at the Nash equilibrium without restrictions, only SP 2's small-cell bandwidth allocation falls below the required threshold. In this case, the new Nash equilibrium is characterized by the following proposition.
\begin{proposition}\label{Prop:One Violated NE}
In case C, the Nash equilibrium with regulatory constraints is one of the following two types:\\
Type I: Both SPs allocate exactly the required minimum amount of bandwidth to small-cells, i.e., $B_{1,S}=B_{1,S}^0, B_{2,S}=B_{2,S}^0$.\\
Type II: Only SP 2 allocates exactly the required minimum amount of bandwidth to small-cells, i.e., $B_{1,S} > B_{1,S}^0, B_{2,S}=B_{2,S}^0$. (See Appendix \ref{Appen:C} for proof.)
\end{proposition}

Note that equation (\ref{Eqn:Boundary Point NE}) also gives conditions when a type I equilibrium arises.

While the type I Nash equilibrium in both cases B and C indicate both SPs allocate exactly the required minimum amount to small-cells, they are quite different. In case B both SPs increase their small-cell bandwidth allocations, whereas in case C, one SP increases its small-cell bandwidth while the other SP decreases its small-cell bandwidth. Another difference is that in case C, the SP whose small-cell bandwidth allocation without restrictions violates the constraint always operates at exactly the required minimum point at the new Nash equilibrium, while it will further increase its small-cell bandwidth beyond the minimum point in a type II equilibrium for case B.
\\

\begin{figure}[htbp]
\centering
\includegraphics[width=0.45\textwidth,height=0.35\textwidth]{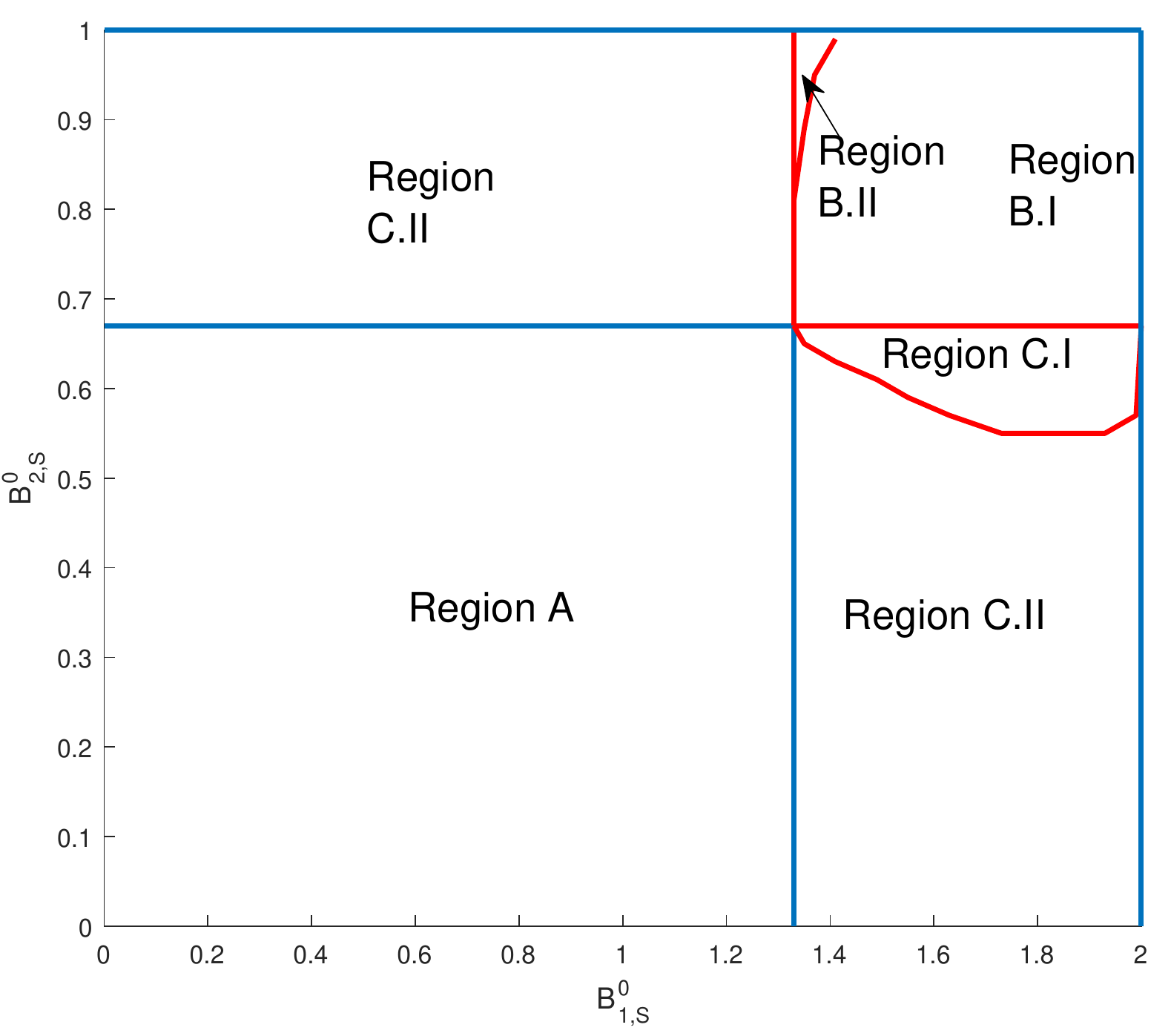}
\caption{Nash equilibrium regions for 2 SPs as the bandwidth restrictions vary.}
\label{Fig:NE_Region}
\end{figure}

Next we use a specific example in Fig. \ref{Fig:NE_Region} to illustrate the different Nash equilibrium regions as a function of the small-cell bandwidth constraints discussed in the preceding cases. The system parameters for this case are: $\alpha=0.5, N_m=N_f=50, R_0=50, \lambda_S=2, B_1=2, B_2=1$. In this example the original equilibrium small-cell bandwidth allocations without the regulatory constraints are: $B_{1,S}=1.34, B_{2,S}=0.67$. Region A corresponds to the Nash equilibrium in case A, which is also the equilibrium without the regulatory constraints. Region B.I and Region B.II correspond to the type-I and type-II Nash equilibrium in case B where both constraints are violated at the original equilibrium, and the same rule applies to Region C.I and C.II.

\section{The Impact of Small-Cell Investment Cost} \label{Sec:Investment}
In this section we study the impacts of small-cell investment on pricing and spectrum allocation in HetNets, with the corresponding optimization problem in (6). Note that in this section we assume no spectrum restrictions are enforced, and as a result SPs are free to split the bandwidth in an arbitrary way. We first assume a single SP and characterize the optimal investment strategy, as well as the corresponding pricing and bandwidth partition across macro- and small-cells. The deployment density largely depends on the per unit deployment cost of small-cells. We then consider a binary investment game in which each SP has the option of investing in a small-cell network with fixed deployment density, and show that equilibria exist in which one, none, or both SPs invest. In addition, there exists asymmetric equilibrium where one SP invests in small-cells while the other doesn't.

\subsection{Monopoly Scenario}
In this section we investigate optimal pricing, investment and bandwidth allocation with a single SP, i.e., the monopoly scenario. Here we assume the SP maximizes revenue, and consider social welfare maximization Section \ref{Sec:SW}.

We first study the pricing decision and bandwidth allocation given fixed investment in small-cells. Once the small-cell deployment density is determined, the investment cost is fixed. As a result, the revenue of the SP only varies with income, i.e., the aggregate amount paid by all users for choosing its macro- or small-cell service. Therefore we can apply the optimal pricing and bandwidth allocation results in our previous work \cite{Opt7-Chen13, Competition5-Chen15}. We have two different service structures depending on whether macro-cells serve fixed users or not:\\
1. {\em Mixed service}: Macro-cells serve both mobile users and a subset of fixed users;\\
2. {\em Separate service}: Macro-cells only serve mobile users.\\
Of course, if $\lambda_S=0$,
then all available bandwidth is assigned
to macro-cells\footnote{Here we drop the SP subscript $i$.}.

\begin{theorem}
\label{Thm:Pricing and Bandwidth Allocation}
Given a fixed small-cell deployment density $\lambda_S$, the optimal prices
and bandwidth allocation for a monopoly SP is determined by the following cases:

1. If $\lambda_S>1$, the optimal bandwidth allocation implies separate service and is given by:
\begin{equation}
B_S^{\text{rev}}=\frac{\epsilon N_fB}{\epsilon N_f+N_m}, \quad B_M^{\text{rev}}=\frac{N_mB}{\epsilon N_f+N_m}
\end{equation}
where $\epsilon=\lambda_S^{\frac{1}{\alpha}-1}$.
Prices are then set to clear the market so that all users are served and all rate is allocated:
\begin{equation}
p_S^{\text{rev}}=\Big( \frac{\epsilon \lambda_S BR_0}{\epsilon N_f+N_m} \Big)^{-\alpha}, \quad p_M^{\text{rev}}=\Big( \frac{BR_0}{\epsilon N_f+N_m} \Big)^{-\alpha}.
\end{equation}

2. If $0\le \lambda_S \le 1$,  all bandwidth is allocated to macro-cells
and it corresponds to the mixed service scenario:
\begin{equation}
B_S^{\text{rev}}=0, \quad B_M^{\text{rev}}=B.
\end{equation}
The optimal prices are
\begin{equation}
p_S^{\text{rev}} =0, \quad p_M^{\text{rev}}=\Big( \frac{BR_0}{N_f+N_m} \Big)^{-\alpha}.
\end{equation}
\end{theorem}

According to Theorem \ref{Thm:Pricing and Bandwidth Allocation},
the optimal prices and bandwidth allocation are uniquely determined
by the investment choice and can be easily calculated.
This will next be used to determine the optimal investment.

From Theorem \ref{Thm:Pricing and Bandwidth Allocation}, we formulate
the corresponding optimization problem in the mixed service scenario as follows:
\begin{align}
\notag \underset{\lambda_S}{\text{maximize}} \quad & S=BR_0u^{\prime}\big(  \frac{BR_0}{N_m+N_f}  \big) -I_S\lambda_S \\
\notag \text{subject to} \quad & 0\le \lambda_S \le 1.
\end{align}

Since in this scenario all bandwidth is allocated to macro-cells,
the optimal small-cell deployment density should be $\lambda_S=0$.
Similarly, when $\lambda_S>1$, corresponding to the separate service scenario, we have:
\begin{align}
\notag \underset{\lambda_S}{\text{max }}  & S=
B_MR_0u^{\prime}
\big(  \frac{B_MR_0}{N_m}  \big)+\lambda_SB_SR_0u^{\prime}
\big(  \frac{\lambda_SB_SR_0}{N_f}  \big) 
-I_S\lambda_S \displaybreak[0] \\
\text{s. t. }  & B_S=\frac{\epsilon N_fB}{\epsilon N_f+N_m}, B_M=\frac{N_mB}{\epsilon N_f+N_m}, \lambda_S>1.
\end{align}
Solving this optimization problem gives the following theorem.

\begin{theorem}[Optimal Investment]\label{Thm:Optimal Investment}
The optimal small-cell deployment density $\lambda_S^{\text{rev}}$
is the maximum of the following two values:
\begin{equation}
\lambda_S^{\text{rev}}=0 \text{ or } \lambda_S^{\text{rev}}=\lambda_S^{*}
\end{equation}
where $\lambda_S^{\text{*}}$ satisfies
\begin{displaymath}
\tag{P1}
\left\{
\begin{array}{ll}
N_f(1-\alpha)(B R_0)^{1-\alpha}{\lambda_S^{*}}^{\frac{1}{\alpha}-2}(N_f{\lambda_S^{*}}^{\frac{1}{\alpha}-1}+N_m )^{\alpha-1}=I_S\\
{\lambda_S^{*}}>1.
\end{array}
\right.
\end{displaymath}
(See Appendix \ref{Appen:D} for proof.)
\end{theorem}

The first profile corresponds to the case that the SP does not invest any amount in small-cells and consequently chooses to allocate all bandwidth to macro-cells. The second profile, however, indicates that the SP would deploy small-cells and allocate some bandwidth in both macro- and small-cells. Note that Theorem \ref{Thm:Optimal Investment} indicates SPs either do not deploy small-cells at all, or deploy the small-cells at a density greater than macro-cells. This conclusion is therefore consistent with the assumption $\lambda_S>1$ in \cite{Opt7-Chen13} and \cite{Competition5-Chen15}.

Depending on the values of $\alpha$ and other system parameters, (P1) may have no solution. In this case the optimal investment choice is  $\lambda_S=0$. The following proposition gives sufficient conditions for this to happen. Specifically, it illustrates that if the per unit deployment cost $I_S$ is large enough, the SP should not invest in small-cells.

\begin{proposition}[Sufficient Conditions for No Investment]\label{Prop:number of solutions}
If the per unit deployment cost $I_S$ exceeds a threshold, (P1) has no feasible solution and therefore the SP should not invest in small-cells.\\
1. When $\alpha\in [\alpha_0,1)$, the threshold is given by:
\begin{equation} \label{Eqn:Inequality1}
I_S\ge N_f(1-\alpha)(BR_0)^{1-\alpha}(N_f+N_m )^{\alpha-1}
\end{equation}
where $\alpha_0$ is the unique solution to the following equation:
\begin{equation}
(N_f+N_m)(1-2\alpha) = N_f (1-\alpha)^2.
\end{equation}
2. When $\alpha\in (0,\alpha_0)$, the threshold is given by:
\begin{equation} \label{Eqn:Inequality2}
I_S\ge N_f{(\lambda_S^0)}^{\frac{1-2\alpha}{\alpha}}(1-\alpha)(BR_0)^{1-\alpha}(N_f{(\lambda_S^0)}^{\frac{1-\alpha}{\alpha}}+N_m )^{\alpha-1}
\end{equation}
where $\lambda_S^0$ is the unique solution to the following equation:
\begin{equation}
\big(N_f+N_m{(\lambda_S^0)}^{1-\frac{1}{\alpha}}\big)(1-2\alpha) = N_f (1-\alpha)^2.
\end{equation}
(See Appendix \ref{Appen:E} for proof.)
\end{proposition}

The intuition behind Proposition \ref{Prop:number of solutions} is as follows. The income of the SP and the deployment cost both increase with the deployment density of small-cells $\lambda_S$ when $\lambda_S>1$. However, if the marginal increase of income with respect to per unit increase of $\lambda_S$ is always smaller than that of investment cost, it would never be beneficial to deploy small-cells for the SP. The investment cost grows linearly with deployment density $\lambda_S$. When $\alpha\in (\alpha_0,1)$, the SP's income is a concave increasing function in $\lambda_S$. When $\alpha\in (\alpha_0,1)$, the income first increases convexly in $\lambda_S$. As $\lambda_S$ further grows, it becomes a concave function in $\lambda_S$ again. Therefore there exists a certain threshold above which deploying small-cells with $\lambda_S>1$ is never beneficial. Moreover, $I_S$ needs to be larger to make the investment in small-cells less attractive.

\begin{figure}[htbp]
\centering
\includegraphics[width=0.45\textwidth,height=0.3\textwidth]{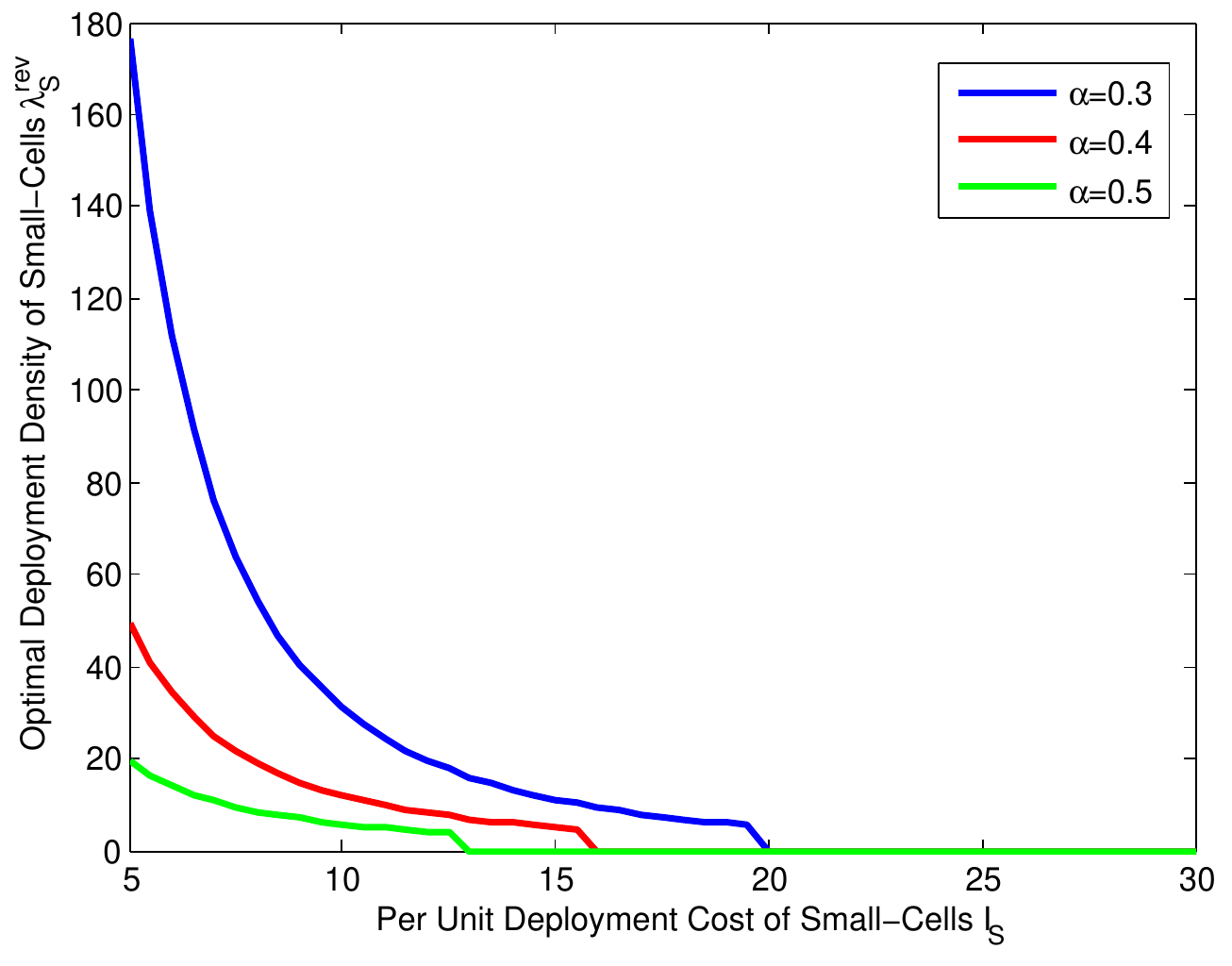}
\caption{Optimal deployment density of small-cells with different per unit deployment cost for a monopoly SP.}
\label{Fig:Optimal Investment}
\end{figure}

Fig.~\ref{Fig:Optimal Investment} illustrates the optimal deployment density of small-cells with the increase of per unit deployment cost. The parameters used are: $R_0=50, N_m=50, N_f=100, B=1$. We can see that as $I_S$ increases, the optimal deployment density of small-cells monotonically decreases until after a certain threshold it reaches zero. Using equations (\ref{Eqn:Inequality1}) and (\ref{Eqn:Inequality2}), we can calculate the sufficient conditions for no investment in small-cells to be  $\alpha=0.5, I_S\ge 28.87, \alpha=0.4, I_S\ge 31.04$, and
$\alpha=0.3, I_S\ge 33.74$, respectively. Note that the figure shows the actual deployent density goes to zero before these conditions apply. This is because even though $(P1)$ has a feasible solution, this solution must be compared with the no investment choice to see which one generates more revenue. Additionally note that smaller values of $\alpha$ result in larger deployment densities of small-cells and so require larger deployment costs before the denisty goes to zero.

\subsection{Competitive Scenario}
We now turn to the competitive scenario with more than one SP. In the monopoly scenario, we considered that the investment decision, $\lambda_S$, was a continuously valued. However, the analysis with this assumption is much harder for the competitive scenario. The primary challenge is that given fixed and continuous investment choices by different SPs, although the price equilibrium remains the same, i.e., SPs should always set the market clearing price, it's very difficult to derive the bandwidth allocation equilibrium analytically. In most cases we can only compute the bandwidth allocation equilibrium numerically, which makes it difficult to get analytical insights into the investment stage. To avoid this difficulty, we simplify the model by making the investment choice of each SP binary, i.e., we assume that each SP can either invest or not in small cells at a given, fixed deployment density, $\lambda_0$. That is, the SPs can only choose between $\lambda_S=0$ and $\lambda_S=\lambda_0$; we refer to this as a binary investment game.   From the previous section we see that if $\lambda_S\le 1$, the SP would allocate all bandwidth to macro-cells since in this case small-cells generate less rate while also incurring investment cost. Thus, we assume $\lambda_0>1$ here. We further focus on a symmetric model, where each SP has the same amount of bandwidth $B$.

For the binary investment game, we have four different cases in terms of investment choices. We next characterize the best response strategies of the two SPs and the corresponding revenue achieved in each case. (See Appendix \ref{Appen:F} for proof.)

\noindent 1. If both SPs choose to invest in small-cells with deployment density $\lambda_0$. Using the results in our previous work \cite{Competition5-Chen15}, both SPs would have the same bandwidth allocation between macro- and small-cells and the corresponding revenues are given by:
\begin{align}
&B_{1,S}=B_{2,S}=\frac{\epsilon_0 N_fB}{\epsilon_0 N_f+N_m}, \displaybreak[0] \\
& B_{1,M}=B_{2,M}=\frac{N_mB}{\epsilon_0 N_f+N_m}, \displaybreak[0]\\
&S_1=S_2=2^{-\alpha}(BR_0)^{1-\alpha}(\epsilon_0 N_f+N_m)^{\alpha}-\lambda_0I_S
\end{align}
where $\epsilon_0=\lambda_0^{\frac{1}{\alpha}-1}$.

\noindent 2. If neither SP invests in small-cells, then the case becomes trivial. The revenue is given by:
\begin{equation}
S_1=S_2=2^{-\alpha}(BR_0)^{1-\alpha}(N_m+N_f)^{\alpha}
\end{equation}

\noindent 3. If SP 1 invests in small-cells while SP 2 doesn't, the bandwidth allocation of SP 1 and the corresponding revenue achieved by two SPs are as follows:

1) If $N_f > \lambda_0 N_m$,
\begin{equation}
B_{1,S}=\min(B_{1,S}^{\text{rev}},B), B_{1,M}=B-B_{1,S}
\end{equation}
\begin{align}
\notag S_1= &\lambda_0B_{1,S}R_0\Big(\frac{\lambda_0B_{1,S}R_0}{N_f}\Big)^{-\alpha}+\allowbreak \\
&B_{1,M}R_0\Big(\frac{(B+B_{1,M})R_0}{N_m}\Big)^{-\alpha}-\lambda_0I_S \displaybreak[0]\\
S_2=&BR_0\Big(\frac{(B+B_{1,M})R_0}{N_m}\Big)^{-\alpha}
\end{align}
where $B_{1,S}^{\text{rev}}$ is the solution to the following equation:
\begin{align}
&\Big(  \frac{(2B-B_{1,S}^{\text{rev}})R_0}{N_m}  \Big)^{-\alpha}+\frac{\alpha BR_0}{(1-\alpha)N_m} \Big(  \frac{(2B-B_{1,S}^{\text{rev}})R_0}{N_m}  \Big)^{-\alpha-1} \displaybreak[0] \\
&=\lambda_0\Big(  \frac{\lambda_0B_{1,S}^{\text{rev}}R_0}{N_f}  \Big)^{-\alpha}
\end{align}

2) If $N_f\le \lambda_0 N_m$,
\begin{align}
&B_{1,S}=B, B_{1,M}=0\\
\notag &S_1= \lambda_0 B R_0\Big(  \frac{(\lambda_0+1)BR_0}{N_m+N_f}  \Big)^{-\alpha}-\lambda_0I_S \displaybreak[0] \\
&S_2=BR_0\Big(  \frac{(\lambda_0+1)BR_0}{N_m+N_f}  \Big)^{-\alpha}
\end{align}

4. Due to symmetry, the case that SP 2 invests in small-cells while SP 1 doesn't is exactly the same as Case 3.

\begin{figure}[htbp]
\centering
\includegraphics[width=0.45\textwidth,height=0.3\textwidth]{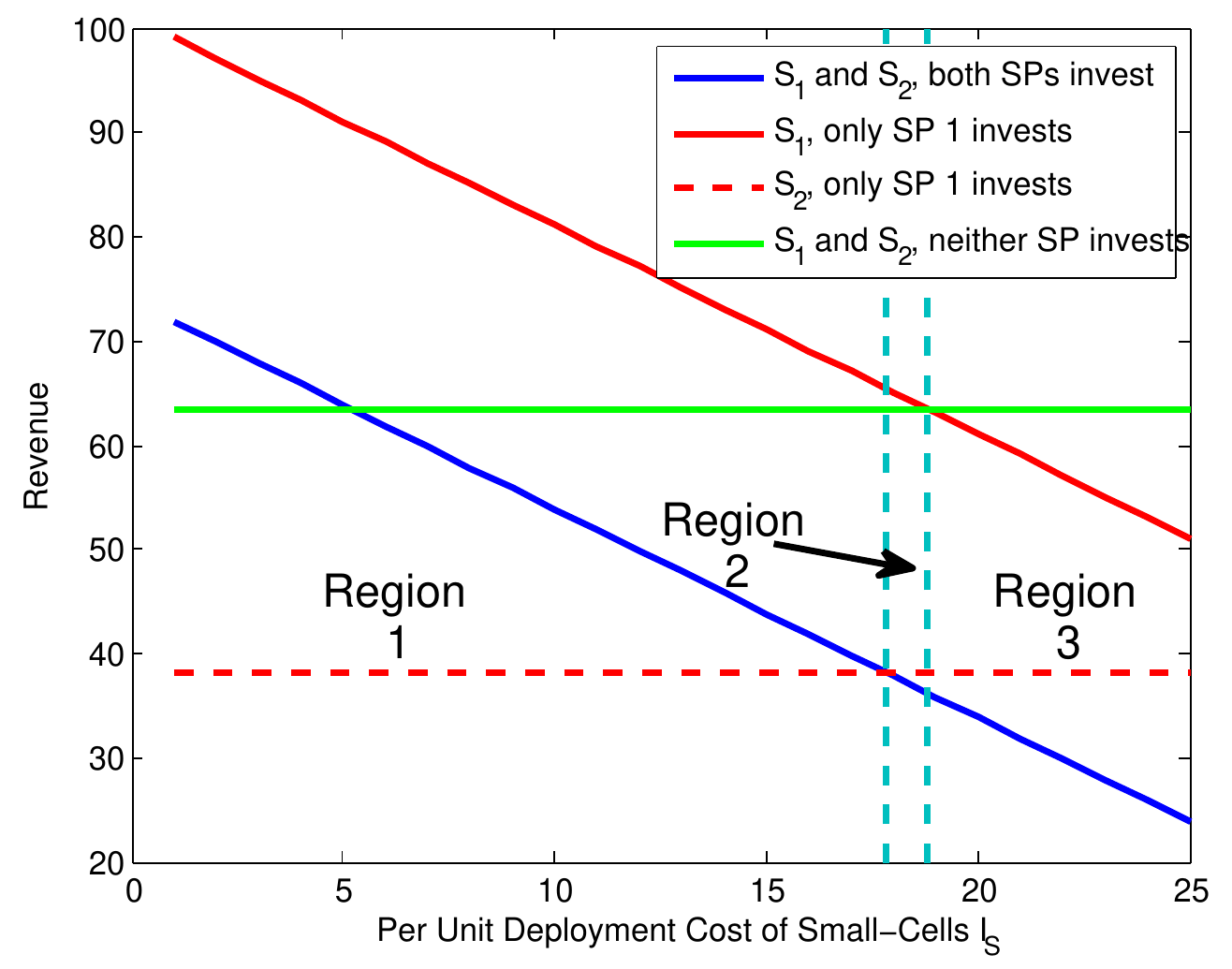}
\caption{Revenue achieved for two revenue-maximizing SPs in the binary investment game.}
\label{Fig:Game}
\end{figure}

We next use a specific example to illustrate the analysis described above. The parameters we use are: $\alpha=0.7, R_0=50, N_m=40, N_f=100, B=1, \lambda_0=2$. Fig. \ref{Fig:Game} shows the revenue achieved by two SPs for the binary investment game when we vary the per-unit deployment cost.

We divide the figure into three regions and it's easy to verify that in Region 1 where the per unit deployment cost is very small, the case that both SPs invest is a Nash equilibrium. In contrast, one SP investing while the other doesn't becomes a Nash equilibrium in Region 2, where the deployment cost is medium. In Region 3, where the deployment cost is very large, the Nash equilibrium is that neither SP invests. This numerical example shows that even for the simple binary investment game, all four investment cases are possible Nash equilibriums, depending on the specific parameters we choose.

In our previous work \cite{Competition5-Chen15}, we showed that if two SPs are symmetric, i.e., if they have the same total amount of bandwidth, at Nash equilibrium their strategies must also be the same. However, when we consider investment, it is possible to have an asymmetric Nash equilibrium even for two symmetric SPs.

\section{Social Welfare Analysis}\label{Sec:SW}
We conduct social welfare analysis in this section, with the intent to evaluate the impact of small-cell spectrum restrictions and investment costs, respectively.  In \cite{Opt7-Chen13}\cite{Competition5-Chen15}, it was shown that for the set of $\alpha$-fair utility functions we use here, the bandwidth allocation at equilibrium is always socially optimal in both monopoly and competitive scenarios. However, the introduction of regulatory bandwidth constraints or deployment cost will change this behavior, and impact the social welfare in a different way.

\subsection{Social Welfare with Small-Cell Spectrum Restrictions}
We start by considering the social welfare problem with spectrum restrictions in (\ref{Opt:SW optimization with regulations}). With a monopoly SP, it is easy to show that the optimal social welfare-maximizing bandwidth allocation strategy is the same as the revenue-maximizing case given in Theorem \ref{Thm:Monopoly Bandwidth Allocation}.
\\

With two competing SPs, in \cite{Opt7-Chen13}\cite{Competition5-Chen15}, we showed that for the set of $\alpha$-fair utility functions we use here, the bandwidth allocation at equilibrium is always socially optimal in both monopoly and competitive scenarios. With the additional regulatory constraints on the minimum amount of small-cell bandwidth allocations, this is not necessarily true. Obviously, if the equilibrium without restrictions already satisfies the regulatory constraints, then the preceding result still holds, i.e., in case A in the previous section. Otherwise, a social welfare loss is incurred compared to the case without regulatory constraints. Denote $\text{SW}_{\text{wo}}^{*}, \text{SW}_{\text{w}}^{\text{NE}}$ as the equilibrium social welfare without and with regulatory constraints, respectively. The following theorem states that the loss in social welfare is lower bounded, and the worst point occurs at the scenario where the regulatory constraints require both SPs to allocate all bandwidth only to small-cells.

\begin{theorem} \label{Thm:SW_Restrictions}
Compared to the case without the regulatory constraints, social welfare loss is incurred when the following inequality holds:
\begin{equation}
\frac{N_f\lambda_S^{1/\alpha-1}}{N_f\lambda_S^{1/\alpha-1}+N_m} \sum\limits_{i\in \mathcal{N}}B_i< \sum\limits_{i\in \mathcal{N}}B_{i,S}^0.
\end{equation}
We have:
\begin{equation}
\frac{\text{SW}_{\text{w}}^{\text{NE}}}{\text{SW}_{\text{wo}}^{*}}
\ge \Big(\frac{N_f\lambda_S^{1/\alpha-1}}{N_f\lambda_S^{1/\alpha-1}+N_m}\Big)^{\alpha},
\end{equation}
where the bound is tight exactly when $B_{i,S}^0=B_i, \forall i\in \mathcal{N}$. (See Appendix \ref{Appen:G} for proof.)
\end{theorem}

In practice, a spectrum regulator, such as the FCC, may seek to find an optimal way to allocate newly available spectrum so that the market equilibrium yields the largest social welfare. We next use our results to analyze the case where the spectrum regulator needs to allocate a total available new bandwidth $B$ to two competitive SPs. SP 1 and 2 each have initial licensed bandwidth $B_{1}^o$ and $B_{2}^o$, and get a proportion of the new bandwidth, denoted as $B_{1}^n$ and $B_2^n$. The initial bandwidth is free to use for either macro-cells or small-cells. In contrast, the new bandwidth can only be used for small-cells. As mentioned before this is motivated by the 3.5GHz band, where FCC regulates the power constraint to be very small, and therefore it can only be used for small-cell deployment \cite{FCC}.

The spectrum regulator needs to determine the optimal split of the new bandwidth such that the social welfare under market equilibrium is maximized. We consider the following three scenarios for any possible bandwidth partition $(B_1^n, B_2^n)$:

1) The optimal social welfare without regulatory constraints, $\text{SW}_{\text{wo}}^{*}$. Note, from \cite{Competition5-Chen15}, this is the same as the equilibrium social welfare without regulatory constraints.
This will be used as a benchmark.

2) The optimal social welfare with the regulatory constraints, which we denote as $\text{SW}_\text{w}^{*}$.

3) The equilibrium social welfare with regulatory constraints, $\text{SW}_{\text{w}}^{\text{NE}}$.

The next theorem compares the three scenarios depending on the total amount of newly available bandwidth $B$.

\begin{theorem}\label{Thm:Three Scenarios}
Given an amount of new bandwidth $B$, there exists a bandwidth threshold $T$
\begin{equation}
T=\frac{(B_1^o+B_2^o)N_f\lambda_S^{1/\alpha-1}}{N_m},
\end{equation}
which determines the following relations:\\
1. If $B>T$, then  $\text{SW}_{\text{w}}^{\text{NE}}\le \text{SW}_{\text{w}}^{*} < \text{SW}_{\text{wo}}^{*}$. The first inequality is binding, i.e., $\text{SW}_{\text{w}}^{\text{NE}}= \text{SW}_{\text{w}}^{*} < \text{SW}_{\text{wo}}^{*}$, if and only if (\ref{Eqn:Boundary Point NE}) holds. \\
2. If $B\le T$, then  $\text{SW}_{\text{w}}^{\text{NE}}\le \text{SW}_{\text{w}}^{*} = \text{SW}_{\text{wo}}^{*}$. The first inequality is binding, i.e., $\text{SW}_{\text{w}}^{\text{NE}} = \text{SW}_{\text{w}}^{*} = \text{SW}_{\text{wo}}^{*}$, if and only if the following condition is met:
\begin{equation}
B_1^n\in \Big[  B-\frac{B_2^oN_f\lambda_S^{1/\alpha-1}}{N_m}, \frac{B_1^oN_f\lambda_S^{1/\alpha-1}}{N_m}  \Big], B_2^n=B-B_1^n.
\end{equation}
(See Appendix \ref{Appen:H} for proof.)
\end{theorem}

Theorem \ref{Thm:Three Scenarios} states that if the total amount of newly available bandwidth is too large, no matter if the two competing SPs maximize revenue or social welfare, we always have some social welfare loss compared to the case without regulatory constraints. This can be explained as follows. Using the set of $\alpha$-fair utility functions, without regulatory constraints the socially optimal bandwidth allocation strategy is to allocate bandwidth to macro- and small-cells based on a fixed proportion. If the total amount of newly available bandwidth is not large, simply following the original allocation satisfies the regulatory requirement and is therefore socially optimal. However, when the amount of new bandwidth becomes large, since the new bandwidth is required to be allocated to small-cells only, the original optimal proportion would violate the small-cell bandwidth constraints. As a result of this, social welfare loss relative to the original allocation scheme becomes inevitable. Further, note that the bandwidth threshold at which this loss occurs is proportional to $\frac{N_f}{N_m}$, so that when there are more fixed users willing to use small-cells, the threshold increases. It is also increasing in $\lambda_S$, the gain in spectral efficiency of small-cells and in the initial allotment of licensed bandwidth.

Theorem \ref{Thm:Three Scenarios} also indicates that when the amount of newly available bandwidth is below the threshold, there exists a bandwidth split that achieves the optimal benchmark social welfare. This result suggests that if a spectrum controller is planning to enforce bandwidth restrictions on newly released bands, it should consider the possible impacts on the market equilibrium. In particular, if the amount of newly available bandwidth is too large, imposing such restrictions might lead to social welfare loss compared to the scenario where the restrictions were not imposed. On the other hand, if the amount of new spectrum is small compared to the existing bands already licensed to SPs in the market, the influence on the market equilibrium from the introduction of bandwidth restrictions on the new bands is minor and controllable, and therefore will not incur any loss in the social welfare.

\begin{figure}[htbp]
\centering
\includegraphics[width=0.45\textwidth,height=0.35\textwidth]{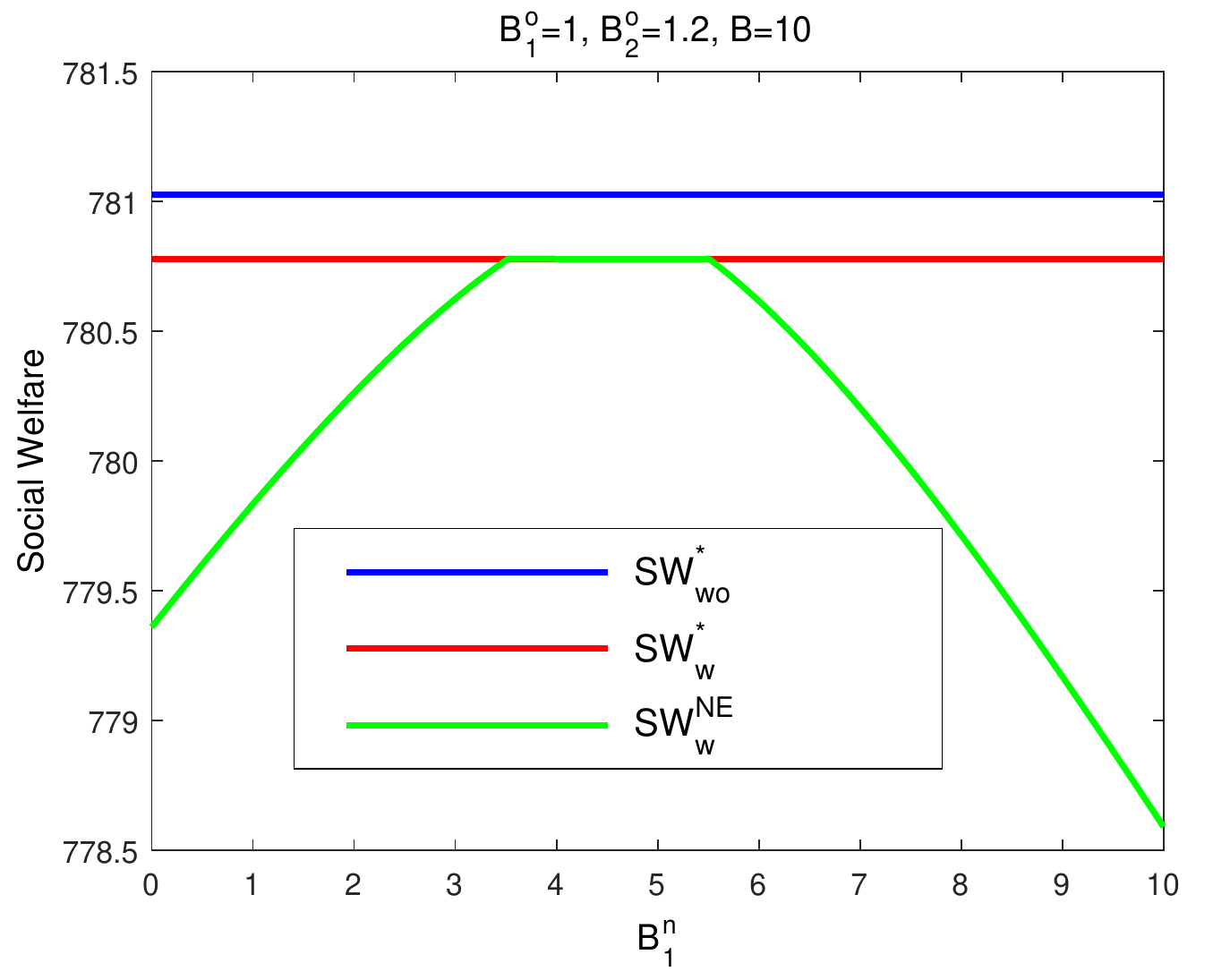}
\caption{Social welfare versus $B_1^n$ with large $B$.}
\label{Fig:SW_2}
\end{figure}

\begin{figure}[htbp]
\centering
\includegraphics[width=0.45\textwidth,height=0.35\textwidth]{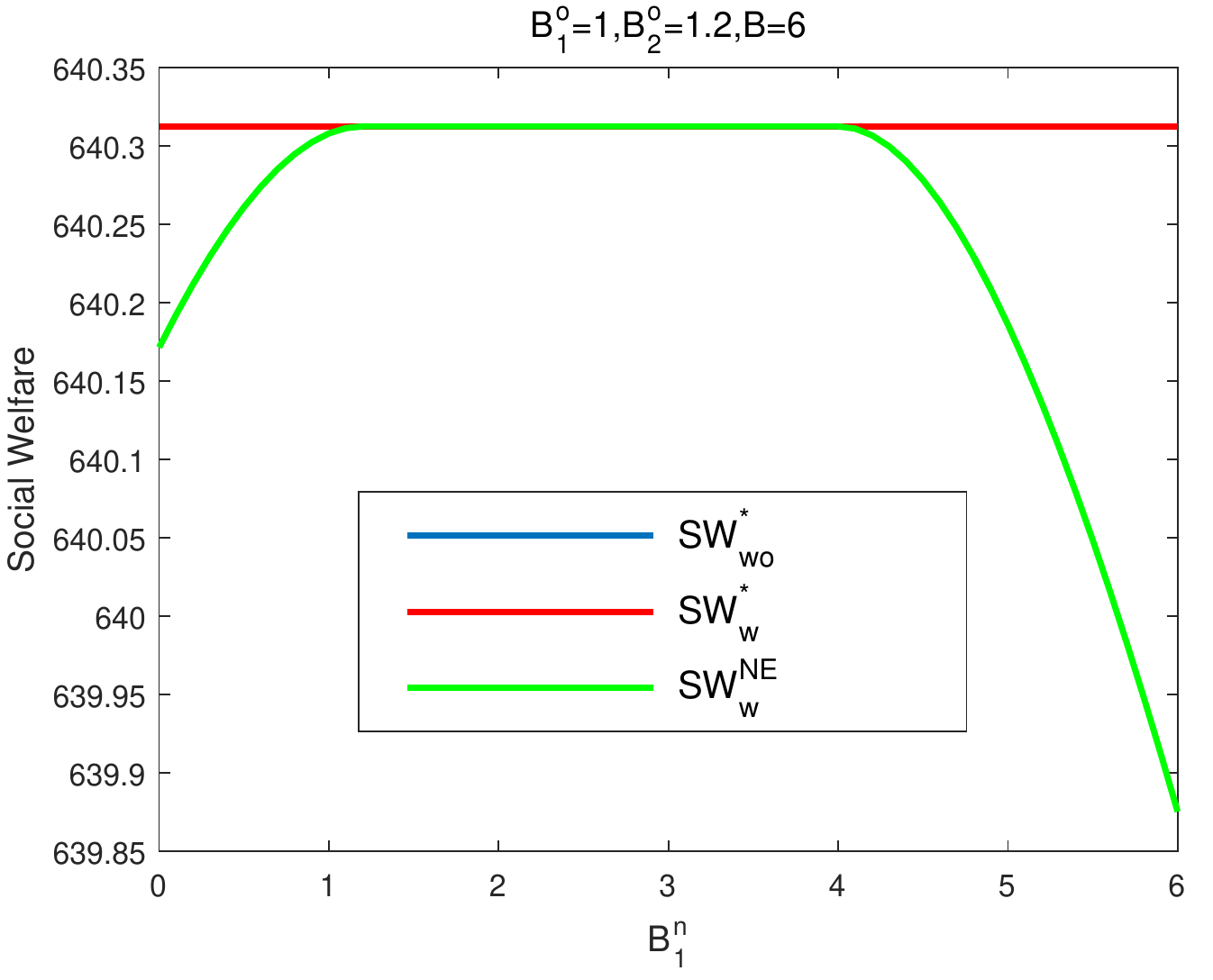}
\caption{Social welfare versus $B_1^n$ with small $B$. }
\label{Fig:SW_1}
\end{figure}

Figures \ref{Fig:SW_2} and \ref{Fig:SW_1} illustrate Theorem \ref{Thm:Three Scenarios}. The system parameters we use in both cases are: $\alpha=0.5, N_m=N_f=50, R_0=50,\lambda_S=4, B_1^o=1, B_2^o=1.2$. The Figures differ in the amount of new bandwidth. In Fig. \ref{Fig:SW_2}, $B=10$, while in Fig. \ref{Fig:SW_1}, $B=6$.  We can see that when the amount of newly available bandwidth is not large, there is a bandwidth split that achieves the optimal benchmark social welfare. However, when the amount of new bandwidth is large relative to the amount of original bandwidth of the SPs, there exists no bandwidth split that achieve the optimal social welfare without the constraints.

\subsection{Social Welfare with Small-Cell Investment Costs}
In this subsection, we turn to the social welfare optimization problem given in (\ref{Opt:SW optimization with investment}). Similarly, we start with  the monopoly scenario. After introducing the deployment cost of small-cells, and changing the SP's objective to maximizing social welfare, the next theorem summarizes the properties of the optimal investment, bandwidth allocation, and pricing.
\begin{theorem}
\label{Thm:SW_Investment}
Given a social welfare maximizing SP, we have the following properties:\\
1) The optimal pricing and bandwidth allocation strategy under fixed deployment density is the same as that of revenue maximization stated in Theorem \ref{Thm:Pricing and Bandwidth Allocation}.\\
2) The optimal deployment density for small-cells is very similar to that of revenue maximization except for an additional factor of $\frac{1}{1-\alpha}$. It can only occur at one of the following two sets of points:
\begin{equation}
\lambda_S^{\text{sw}}=0 \text{ or } \lambda_S^{\text{sw}}=\lambda_S^{*}
\end{equation}
where $\lambda_S^{*}$ is the solution to the following equation:
\begin{displaymath}
\tag{P1}
\left\{
\begin{array}{ll}
N_f(BR_0)^{1-\alpha}{\lambda_S^{*}}^{\frac{1}{\alpha}-2}(N_f{\lambda_S^{*}}^{\frac{1}{\alpha}-1}+N_m )^{\alpha-1}=I_S,\\
{\lambda_S^{*}}>1.
\end{array}
\right.
\end{displaymath}
For the same set of parameters, if both $\lambda_S^{\text{sw}}>1$ and $\lambda_S^{\text{rev}}>1$, then $\lambda_S^{\text{sw}}>\lambda_S^{\text{rev}}$.
\end{theorem}

Theorem~\ref{Thm:SW_Investment} shows that compared with revenue maximization, if the SP operates in small-cells (when $\lambda_S>0$), social welfare maximization requires the SP to make a larger investment in small-cells. This observation is easily explained once we take a deeper look at the market structure in our model. The social welfare is the sum utility of all users minus the investment cost, while the revenue is defined as the income of the SP less the investment cost. In equation (\ref{Eqn:User Net Payoff}), $u(D(p))$ is the utility and $pD(p)$ is the revenue of the SP. For $\alpha$-fair utility functions, we can easily verify that $pD(p)=(1-\alpha)U(D(p))$. For both revenue maximization and social welfare maximization we need to subtract the same investment cost. However, with revenue maximization the income part is only a fraction of the sum utility. As a result, the marginal income increase with respect to an increase in $\lambda_S$ is always smaller than that of the marginal utility increase, which leads to the result that revenue maximization has a  smaller deployment density in small-cells. \\

\begin{figure}[htbp]
\centering
\includegraphics[width=0.45\textwidth,height=0.3\textwidth]{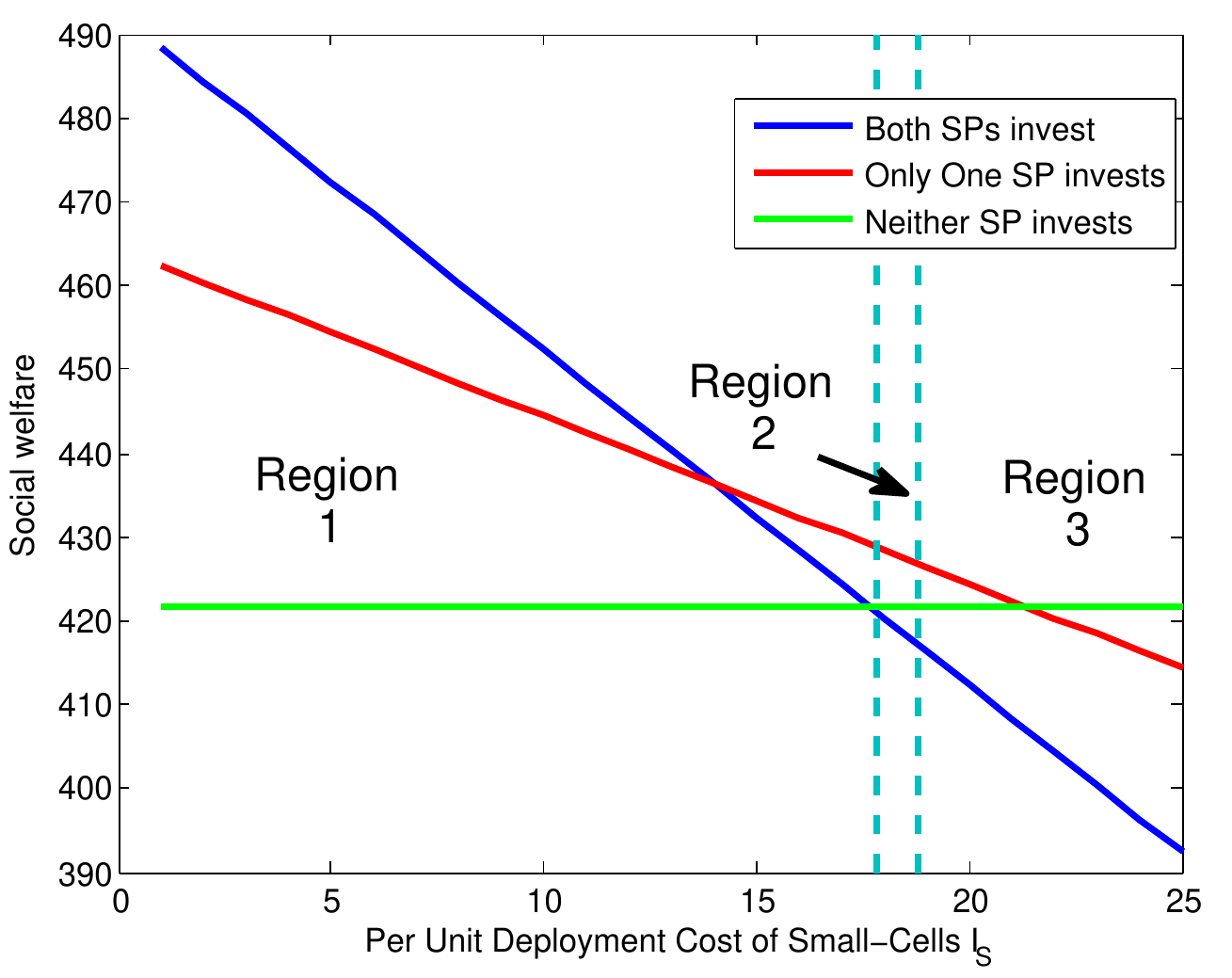}
\caption{Social welfare achieved for two revenue-maximizing SPs in the binary investment game.}
\label{Fig:Game_SW}
\end{figure}

We next see the social welfare performance of different Nash equilibria of the binary investment game introduced in Section \ref{Sec:Investment}. Fig. \ref{Fig:Game_SW} shows the social welfare achieved if SPs perform revenue-maximizing strategies with the same parameters as in Fig. \ref{Fig:Game}.
We can see that while that both SPs investing in small-cells should be the Nash equilibrium in Region 1, the social welfare achieved may be less than that from the case only one SP invests. On the other hand, in Region 3 the Nash equilibrium is that no SP invests in small-cells, whereas the social welfare associated with the case that only one SP invests is larger within a range of $I_S$. From this example we conclude that the Nash equilibria of the binary investment game are not necessarily socially optimal. This is significantly different from the result in \cite{Competition5-Chen15} where the Nash equilibria corresponding to $\alpha$-fair utility functions with fixed deployment density of small-cells are always socially optimal.

\begin{figure}[htbp]
\centering
\includegraphics[width=0.45\textwidth,height=0.3\textwidth]{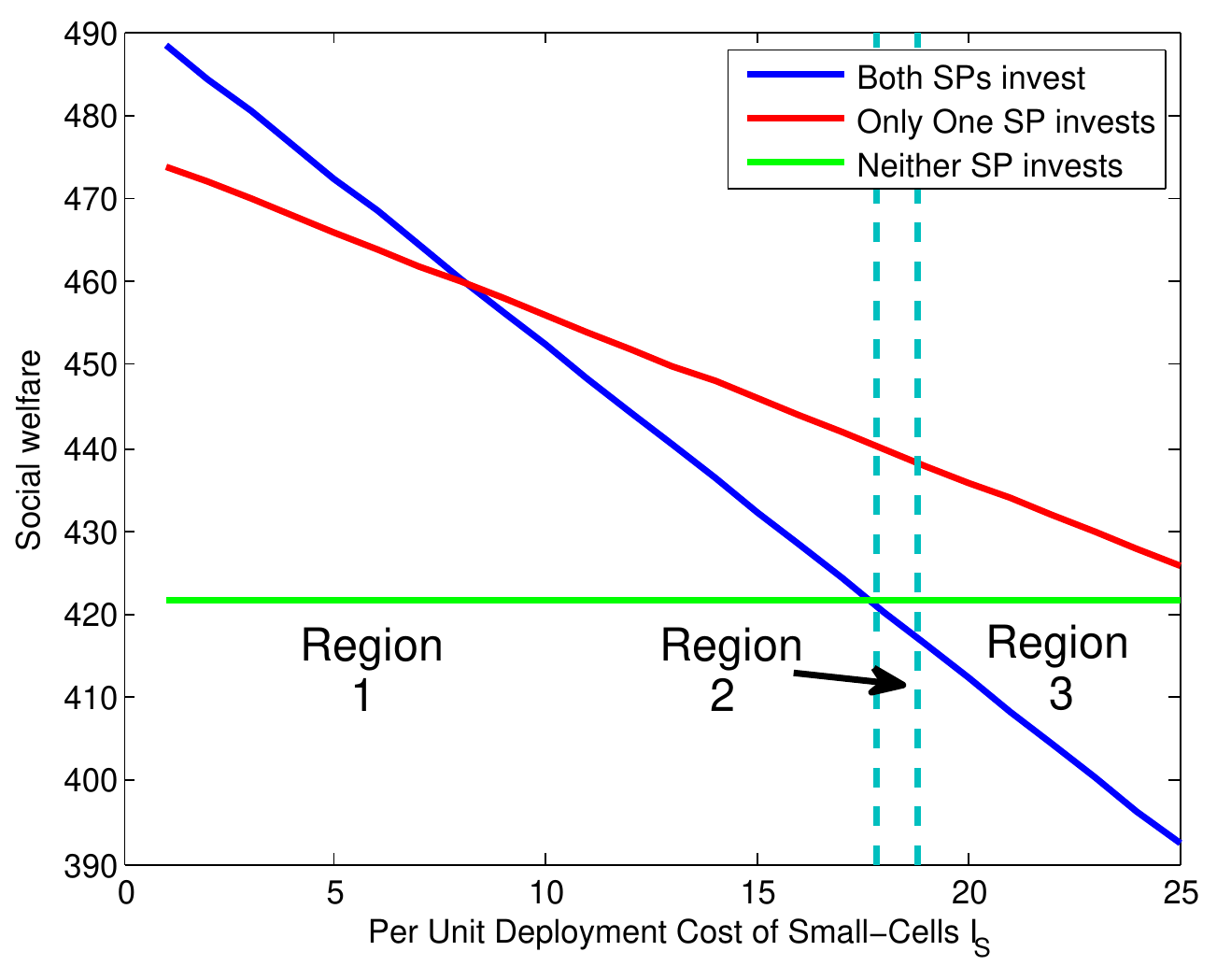}
\caption{Social welfare achieved for two social welfare-maximizing SPs in the binary investment game.}
\label{Fig:Game_SW_2}
\end{figure}

If we assume the SPs perform social welfare-maximizing strategies, the bandwidth allocation when two SPs both invest would be the same as revenue-maximizing SPs case \cite{Competition5-Chen15}. Consequently, the social welfare would be the same when either both SPs invest or neither SP invests. When only one SP invests, the bandwidth allocation is different and it can be proved that in this case social welfare-maximizing SPs would allocate more bandwidth to small-cells compared to revenue-maximizing SPs. Fig. \ref{Fig:Game_SW_2} illustrates the social welfare achieved if SPs use such strategies with the same setting as in Fig. \ref{Fig:Game_SW}. Fig. \ref{Fig:Game_SW_2} shows that the social welfare corresponding to the case where only one SP invests is now larger.

\section{Conclusions}\label{Sec:Conclusions}
In this paper we considered the impact of spectrum restrictions and investment costs on the pricing and bandwidth allocation decisions by both a monopolist and competitive SPs. Moreover, we also evaluated the corresponding social welfare implications of these factors.

By imposing a required minimum bandwidth allocation on small-cells, the optimal bandwidth allocation strategies of SPs can change dramatically from the unrestricted case. While this change is relatively straightforward in the monopoly scenario, it turns out to be much more complicated in the competitive scenario with two SPs. Specifically, the existence and uniqueness of Nash equilibria are still preserved after adding these constraints. However, the equilibria can exhibit very different structures and characteristics as the constraints vary. We showed that the introduction of such spectrum constraints may shift the equilibrium away from an efficient allocation, thus incurring some social welfare loss.

In contrast, by considering the deployment cost of small-cells, we showed that for a monopolist SP, the optimal deployment density of small-cells largely depends on the per unit deployment cost. If the per unit deployment cost is large enough, a revenue-oriented SP may decide not to deploy small-cells. If social welfare is the objective, and if the SP does invest in small-cells, then it will deploy a higher density of small-cells than when maximizing revenue. With multiple SPs, we considered a simplified binary investment game and showed that different types of Nash equilibrium are possible as the unit deployment cost varies. Depending on the investment cost, at Nash equilibrium, the number of SPs that invest in small-cells can be zero, one or two. However, the Nash equilibria may not be socially optimal.

Our results show that both spectrum restrictions and investment costs
can result in some welfare loss compared to the case where they are absent. To avoid this loss, a regulator could attempt to avoid restricting spectrum usage and try to lower the investment costs in small cells. Of course, there will be other considerations that regulators need to account for such as the sharing constraints in the 3.5 GhZ band that may motivate not adopting these suggestions.

For future directions, instead of studying small-cell spectrum restrictions and investment costs individually, one could jointly consider these effects. Another possible direction is to consider the addition of unlicensed spectrum in which multiple providers could deploy small cells using techniques such as WiFi, LTE-U, or LAA as in \cite{Unlicensed1-Chen16}. It would also be useful to study the impact of other policy and operational decisions, and also a broader set of applications and radio access technologies.

\ifCLASSOPTIONcaptionsoff
  \newpage
\fi

\nocite{*}
\bibliography{References}{}
\bibliographystyle{ieeetr}

%

%
%
%




\begin{appendices}
\section{Proof of Theorem \ref{Thm:Monopoly Bandwidth Allocation}} \label{Appen:A}
The proof of Theorem \ref{Thm:Monopoly Bandwidth Allocation} is straightforward and we can apply the results in \cite{Opt7-Chen13} directly. In particular, if the original optimal bandwidth allocation still holds with the added restriction constraints, then we are done. Otherwise we have to increase the small-cell bandwidth allocation. Both the revenue and social welfare are concave functions in the small-cell bandwidth allocation, and at the original equilibrium point the marginal revenue and social welfare increase with respect to per unit increase in bandwidth are equal for both macro- and small-cells. Hence, when the small-cell bandwidth increases, we enter the region where the marginal revenue and social welfare increase with respect to per unit increase in bandwidth for small-cells is smaller than that of macro-cells. As a result, the best option is to operate at the boundary point, i.e., allocating exactly the required minimum amount of bandwidth to small-cells.

\section{Proof of Theorem \ref{Thm:NE}} \label{Appen:B}
As this is a concave game, to prove the existence and uniqueness of the Nash equilibrium, we can use the uniqueness theorem (Theorem 6) in Rosen's paper \cite{Rosen65}, which gives sufficient condition in terms of a certain matrix being negative definite. In our previous work \cite{Competition5-Chen15} it was proved that the required matrix is negative definite for the corresponding game without bandwidth restrictions. Here the only difference is that we have additional linear constraints on the bandwidth allocations, which do not have any effect on this result. Therefore, the same arguments also apply here.

As for the second part of the theorem, denote $R_S$ and $R_S^{\prime}$ as the average service rate in small-cells with and without the regulatory constraints, respectively. Suppose at the Nash equilibrium with constraints, the sum bandwidth allocation to small-cells is less than that without the regulatory constraints, then we have:
\begin{equation}
R_S^{\prime}<R_S.
\end{equation}

Denote $D_i=\frac{\partial S_i}{\partial B_{i,S}}$, it follows that:
\begin{align}
\nonumber  D_1+D_2=&\lambda_S\Big[2u^{\prime}(R_S)+R_Su^{\prime\prime}(R_S)\Big]-\\
&\Big[2u^{\prime}(R_M)+R_Mu^{\prime\prime}(R_M)\Big].
\end{align}

Since $2u^{\prime}(r)+ru^{\prime\prime}(r)$ decreases in $r$, and we know that at the Nash equilibrium without constraints, $D_1=D_2=0$, we can conclude that $D_1+D_2>0$ at the equilibrium with constraints. As a result, at least one of $D_1$ or $D_2$ must be greater than 0 at equilibrium. Without loss of generality, suppose $D_2>0$ at the equilibrium with constraints.

Given $D_2>0$ , it must be that $B_{2,S}^{\prime}=B_2$, and $D_1\le 0$. This is because if $D_1>0$ also holds, $B_{1,S}^{\prime}=B_1$ and it contradicts with the fact that $R_S^{\prime}<R_S$.

Then at the Nash equilibrium without constraints, we have:
\begin{align}
\nonumber D_1=&\lambda_Su^{\prime}(R_S)+\lambda_S^2\frac{B_{1,S}R_0}{N_f}u^{\prime\prime}(R_S)\\
&-u^{\prime}(R_M)-\frac{B_{1,M}R_0}{N_m}u^{\prime\prime}(R_M)\\
\nonumber =&\lambda_S\Big[u^{\prime}(R_S)+R_Su^{\prime\prime}(R_S)\Big]-\Big[u^{\prime}(R_M)+R_Mu^{\prime\prime}(R_M)\Big]\\
& -\lambda_S^2\frac{B_{2,S}R_0}{N_f}u^{\prime\prime}(R_S)+\frac{B_{2,M}R_0}{N_m}u^{\prime\prime}(R_M)=0. \label{Appen_B_Inequality1}
\end{align}

At the equilibrium with constraints, similarly we have:
\begin{align}\label{Appen_B_Inequality2}
\nonumber D_1=& \lambda_S\Big[u^{\prime}(R_S^{\prime})+R_S^{\prime}u^{\prime\prime}(R_S^{\prime})\Big]-
\Big[u^{\prime}(R_M^{\prime})+R_M^{\prime}u^{\prime\prime}(R_M^{\prime})\Big]\\
& -\lambda_S^2\frac{B_2R_0}{N_f}u^{\prime\prime}(R_S^{\prime})\le 0.
\end{align}

However, since $u^{\prime}(r)+ru^{\prime\prime}(r)$ decreases in $r$, $u^{\prime\prime}(r)<0$ and increases in $r$, and the fact that $R_S^{\prime}<R_S, R_M^{\prime}>R_M$, the inequality sign in (\ref{Appen_B_Inequality2}) should be reversed. Therefore we have a contradiction.

\section{Proof of Proposition \ref{Prop:One Violated NE} \label{Appen:C}}
We prove why it is not possible to have $B_{1,S} = B_{1,S}^0, B_{2,S}>B_{2,S}^0$ at the NE of type II in case C.

If at the new NE with constraints, we have $B_{1,S} = B_{1,S}^0, B_{2,S}>B_{2,S}^0$. We must have $D_1\le 0, D_2\ge 0$. For the latter, we have:
\begin{align}
\notag D_2=& \lambda_S\Big[u^{\prime}(R_S^{\prime})+R_S^{\prime}u^{\prime\prime}(R_S^{\prime})\Big]-
\Big[u^{\prime}(R_M^{\prime})+R_M^{\prime}u^{\prime\prime}(R_M^{\prime})\Big] \\ &-\lambda_S^2\frac{B_{1,S}R_0}{N_f}u^{\prime\prime}(R_S^{\prime})+\frac{B_{2,M}R_0}{N_m}u^{\prime\prime}(R_M)\ge 0.
\end{align}

Without constraints, we have $D_1^{\prime}=0, D_2^{\prime}=0$, and denote the corresponding bandwidth allocation as $(B_{1,S}^{\prime},B_{1,M}^{\prime})$ and $(B_{2,S}^{\prime},B_{2,M}^{\prime})$, we have:
\begin{align}
\notag D_2^{\prime}=& \lambda_S\Big[u^{\prime}(R_S^{\prime})+R_S^{\prime}u^{\prime\prime}(R_S^{\prime})\Big]-
\Big[u^{\prime}(R_M^{\prime})+R_M^{\prime}u^{\prime\prime}(R_M^{\prime})\Big] \\ &-\lambda_S^2\frac{B_{1,S}^{\prime}R_0}{N_f}u^{\prime\prime}(R_S^{\prime})+\frac{B_{2,M}^{\prime}R_0}{N_m}u^{\prime\prime}(R_M)= 0.
\end{align}

However, since $u^{\prime}(r)+ru^{\prime\prime}(r)$ decreases in $r$, $u^{\prime\prime}(r)<0$ and increases in $r$, along with the fact that $B_{1,S}\le B_{1,S}^{\prime}$, $B_{1,M}\ge B_{1,M}^{\prime}$, $R_S> R_S^{\prime}$, and $R_M< R_S^{\prime}$, we can conclude $D_2<D_2^{\prime}=0$. We therefore have a contradiction.

\section{Proof of Theorem \ref{Thm:Optimal Investment}} \label{Appen:D}
When $\lambda_S>1$, we can rewrite the revenue of the SP as follows:
\begin{equation}
S=N_m R_M u^{\prime}(R_M)+N_f R_S u^{\prime}(R_S)-I_S\lambda_S,
\end{equation}
where $R_M,R_S$ are the average rate in macro- and small-cells, respectively. By Theorem \ref{Thm:Pricing and Bandwidth Allocation} we can calculate:
\begin{equation}
R_M=\frac{BR_0}{N_m+\epsilon N_f}, R_S=\frac{\lambda_S^{\frac{1}{\alpha}}BR_0}{N_m+\epsilon N_f}.
\end{equation}

Hence we can express $S$ as a function of $\lambda_S$ and the derivative with respect to $\lambda_S$ is given by:
\begin{align}
&S=(BR_0)^{1-\alpha}(N_m+\epsilon N_f)^{\alpha}-I_S\lambda_S,\\
&\frac{\partial S}{\partial \lambda_S}=N_f(1-\alpha)(BR_0)^{1-\alpha}\lambda_S^{\frac{1}{\alpha}-2}(N_m+\epsilon N_f)^{\alpha-1}-I_S.
\end{align}

In order to find the maximal point, we can set the derivative to zero and find the stationary points.

\section{Proof of Proposition \ref{Prop:number of solutions}} \label{Appen:E}
Calculating the second derivative of $S$ with respect to $\lambda_S$, we have:
\begin{align}
\notag \frac{\partial^{2}S}{\partial \lambda_S^2}=&N_f(1-\alpha)(BR_0)^{1-\alpha}
\Big[
\frac{1-2\alpha}{\alpha}(N_m+\epsilon N_f)^{\alpha-1}\lambda_S^{\frac{1}{\alpha}-3}\\
&-\frac{(1-\alpha)^2}{\alpha}N_f(N_m+\epsilon N_f)^{\alpha-2}\lambda_S^{\frac{2}{\alpha}-4}
\Big]. \label{Eqn:Second derivative}
\end{align}

It's easy to see if $\alpha\in (\frac{1}{2},1)$, the second derivative is always negative. As a result, $S$ is concave with $\lambda_S$.

Moreover, if $\alpha\in (0,\frac{1}{2})$  we do the following notation:
\begin{align}
&P=\frac{1-2\alpha}{\alpha}(N_m+\epsilon N_f)^{\alpha-1}\lambda_S^{\frac{1}{\alpha}-3},\\
&N=\frac{(1-\alpha)^2}{\alpha}N_f(N_m+\epsilon N_f)^{\alpha-2}\lambda_S^{\frac{2}{\alpha}-4}.
\end{align}

$P$ and $N$ denote the absolute value of the positive and negative term in (\ref{Eqn:Second derivative}), respectively.

\begin{equation}
\frac{P}{N}=\frac{(1-2\alpha)(N_m\lambda_S^{1-\frac{1}{\alpha}}+N_f)}{(1-\alpha)^2N_f},
\end{equation}
which clearly decreases with $\lambda_S$. As a result, if $\frac{P}{N}$ is no more than one when $\lambda_S=1$, we can conclude that $S$ is always concave with $\lambda_S$. This is satisfied when:
\begin{equation}
\frac{(1-2\alpha)}{(1-\alpha)^2}\le \frac{N_f}{(N_m+N_f)}.
\end{equation}

Since the left hand side is decreasing with $\alpha$, this condition is simply $\alpha>\alpha_0$ where $\alpha_0$ is the solution to:
\begin{equation}
\frac{(1-2\alpha)}{(1-\alpha)^2}= \frac{N_f}{(N_m+N_f)}.
\end{equation}

When $S$ is concave with $\lambda_S$, if the derivative is less than zero at the initial point $\lambda_S=1$, then we can conclude (P1) has no solution and the SP should not invest in small-cells. This condition is given by:
\begin{equation}
I_S\ge N_f(1-\alpha)(BR_0)^{1-\alpha}(N_f+N_m )^{\alpha-1}
\end{equation}

When $\alpha\in (0,\alpha_0)$, it's easy to see that when $\lambda_S$ is small, $S$ is convex with $\lambda_S$ and when $\lambda_S$ is large, $S$ is concave with $\lambda_S$ again. As a result, the maximum derivative occurs at the turning point when $\frac{P}{N}=1$, i.e.,
\begin{equation}
\frac{1-2\alpha}{(1-\alpha)^2}=\frac{N_f}{N_f+N_m{\lambda_S^0}^{1-\frac{1}{\alpha}}}
\end{equation}

If the derivative at $\lambda_S=\lambda_S^0$ is less than zero, then (P1) has no solution and the SP should not invest in small-cells. This condition is given by:
\begin{equation}
I_S\ge N_f{\lambda_S^0}^{\frac{1}{\alpha}-2}(1-\alpha)(BR_0)^{1-\alpha}(N_f{\lambda_S^0}^{\frac{1}{\alpha}-1}+N_m )^{\alpha-1}
\end{equation}

\section{Proof of results about the binary investment game in Section \ref{Sec:Investment}} \label{Appen:F}
Since the cases that both SPs invest and neither SP invests are trivial, we only prove the optimal bandwidth allocation when only SP 1 invests.

The boundary point between mixed and separate service scenario is $R_S=R_M$, i.e., $\frac{(B+B_{1,M})R_0}{N_m}=\frac{\lambda_0B_{1,S}R_0}{N_f} $. Thus, we can easily verify that if $N_f\le \lambda_0 N_m$, it would always be mixed service scenario.

At mixed service scenario, the revenue of SP 1 is given by:
\begin{align}
\notag S_1=&(B_{1,M}+\lambda_0B_{1,S})R_0u^{\prime}\Big(  \frac{(B+B_{1,M}+\lambda_0B_{1,S})R_0}{N_m+N_f}  \Big)\\
=&R(N_m+N_f)u^{\prime}(R)-\frac{BR_0}{N_m+N_f}u^{\prime}(R)
\end{align}
where $R$ is the average rate in both macro- and small-cells.

Since $Ru^{\prime}(R)$ is increasing with $R$ while $u^{\prime}(R)$ is decreasing with $R$, SP 1 would always have the incentive to increase $B_{1,S}$ at mixed service scenario. When $N_f\le \lambda_0 N_m$, this means $B_{1,S}=B$. When $N_f> \lambda_0 N_m$, this means the optimal would occur at separate service.

At separate service scenario, the revenue of SP 1 is given by:
\begin{equation}
S_1=B_{1,M}R_0u^{\prime}\Big(  \frac{(B+B_{1,M})R_0}{N_m}  \Big)+\lambda_0B_{1,S}R_0u^{\prime}\Big(  \frac{\lambda_0B_{1,S}R_0}{N_f}  \Big)
\end{equation}

Taking the derivative of $S_1$ with respect to $B_{1,M}$ and $B_{1,S}$, respectively, we have:
\begin{align}
\frac{\partial S_1}{\partial B_{1,M}}&=R_0\Big[  u^{\prime}(R_M)+R_Mu^{\prime\prime}(R_M)-\frac{BR_0}{N_m}u^{\prime\prime}(R_M)  \Big]\\
\frac{\partial S_1}{\partial B_{1,S}}&=(1-\alpha)R_0\Big[  u^{\prime}(R_S)+R_Su^{\prime\prime}(R_S)  \Big]
\end{align}

It's easy to see $S_1$ are both concave increasing with $B_{1,M}$ and $B_{1,S}$, therefore the optimal point should occur at the point where the two marginal increases are equal. That is given by:
\begin{align}
\notag &\Big(  \frac{(2B-B_{1,S}^{\text{rev}})R_0}{N_m}  \Big)^{-\alpha}+\frac{\alpha BR_0}{(1-\alpha)N_m} \Big(  \frac{(2B-B_{1,S}^{\text{rev}})R_0}{N_m}  \Big)^{-\alpha-1}\\
&=\lambda_0\Big(  \frac{\lambda_0B_{1,S}^{\text{rev}}R_0}{N_f}  \Big)^{-\alpha}
\end{align}

\section{Proof of Theorem \ref{Thm:SW_Restrictions}} \label{Appen:G}
Applying the same arguments we used in proving Theorem \ref{Thm:Monopoly Bandwidth Allocation}, we know that since increasing the small-cell bandwidth allocation beyond the original equilibrium point only decreases the social welfare, then the worst case occurs at the point that all bandwidth is required to be allocated to small-cells.

\section{Proof of Theorem \ref{Thm:Three Scenarios}} \label{Appen:H}
For scenario 2) and 3), as long as the sum of the small-cell bandwidth allocations of the two SPs at the equilibrium without the constraints is larger than the sum of the restriction constraints, then they are the same. This requires:
\begin{equation}
\frac{N_f\lambda_S^{1/\alpha-1}(B_1^o+B_1^n+B_2^o+B_2^n)}{N_f\lambda_S^{1/\alpha-1}+N_m}\ge B_1^n+B_2^n,
\end{equation}
which yields the following condition:
\begin{equation}
B\le \frac{(B_1^o+B_2^o)N_f\lambda_S^{1/\alpha-1}}{N_m}.
\end{equation}
Otherwise, if the preceding condition is not satisfied, the social welfare corresponding to the second scenario is also less than that corresponding to the first scenario, i.e., $\text{SW}_{\text{w}}^{*} < \text{SW}_{\text{wo}}^{*}$.

On the other hand, the only possible way for scenario 3) to achieve the optimal social welfare corresponding to scenario 1) is to ensure the Nash equilibrium is exactly the same as the one without the restriction constraints. This requires:
\begin{equation}
\frac{N_f\lambda_S^{1/\alpha-1}(B_1^o+B_1^n)}{N_f\lambda_S^{1/\alpha-1}+N_m}\ge B_1^n,
\frac{N_f\lambda_S^{1/\alpha-1}(B_2^o+B_2^n)}{N_f\lambda_S^{1/\alpha-1}+N_m}\ge B_2^n,
\end{equation}
which can be simplified to:

\begin{subequations}
\begin{align}
&B\le \frac{(B_1^o+B_2^o)N_f\lambda_S^{1/\alpha-1}}{N_m},\\
&B_1^n\in \Big[  B-\frac{B_2^oN_f\lambda_S^{1/\alpha-1}}{N_m}, \frac{B_1^oN_f\lambda_S^{1/\alpha-1}}{N_m}  \Big].
\end{align}
\end{subequations}

When $\text{SW}_{\text{w}}^{*} < \text{SW}_{\text{wo}}^{*}$, it means the required minimum sum bandwidth allocation to small-cells is larger than the sum bandwidth in small-cells at the equilibrium without constraints. Since we know that at the original equilibrium the social welfare is maximized and the social welfare is a concave function with respect to the sum bandwidth in small-cells, in this case the social welfare maximizing point with the constraints is therefore exactly the required minimum small-cell bandwidth point, i.e., when $B_{1,S}+B_{2,S}=B_{1,S}^0+B_{2,S}^0$. The only possibility for scenario 3) to achieve this is to ensure $B_{1,S}=B_{1,S}^0, B_{2,S}=B_{2,S}^0$ at the Nash equilibrium with constraints. As a result, equation (\ref{Eqn:Boundary Point NE}) becomes exactly the condition for $\text{SW}_{\text{w}}^{\text{NE}}=\text{SW}_{\text{w}}^{*} $.

\end{appendices}

\end{document}